\documentclass[manuscript]{acmart}

\AtBeginDocument{%
  }

\copyrightyear{2025}
\acmJournal{CSUR}
\acmVolume{?}
\acmNumber{?}
\acmArticle{?}
\acmYear{2025}
\acmDOI{XXXXXXX.XXXXXXX}

\usepackage{svg}
\usepackage{float}
\usepackage{url} 
\usepackage{array} 
\usepackage{multicol}
\usepackage{longtable}

\begin{document}
\svgpath{{./Images/SVG}}
\title{A Survey on Multimodal Music Emotion Recognition}

% \affiliation{%

\author{Rashini Liyanarachchi}
\affiliation{\institution{University of New South Wales}\city{Sydney}\country{Australia}}
\email{r.liyanarachchi_lekamlage@unsw.edu.au}
\orcid{0000-0003-0228-6658}
\author{Aditya Joshi}
\affiliation{\institution{University of New South Wales}\city{Sydney}\country{Australia}}
\email{aditya.joshi@unsw.edu.au}
\orcid{0000-0003-2200-9703}
\author{Erik Meijering}
\affiliation{\institution{University of New South Wales}\city{Sydney}\country{Australia}}
\email{erik.meijering@unsw.edu.au}
\orcid{0000-0001-8015-8358}

\renewcommand{\shortauthors}{Rashini et al.}

\begin{abstract}
Multimodal music emotion recognition (MMER) is an emerging discipline in music information retrieval that has experienced a surge in interest in recent years. This survey provides a comprehensive overview of the current state-of-the-art in MMER. Discussing the different approaches and techniques used in this field, the paper introduces a four-stage MMER framework, including multimodal data selection, feature extraction, feature processing, and final emotion prediction. The survey further reveals significant advancements in deep learning methods and the increasing importance of feature fusion techniques. Despite these advancements, challenges such as the need for large annotated datasets, datasets with more modalities, and real-time processing capabilities remain. This paper also contributes to the field by identifying critical gaps in current research and suggesting potential directions for future research. The gaps underscore the importance of developing robust, scalable, and interpretable models for MMER, with implications for applications in music recommendation systems, therapeutic tools, and entertainment.
\end{abstract}


 

%\ccsdesc[500]{Do Not Use This Code~Generate the Correct Terms for Your Paper}
%\ccsdesc[300]{Do Not Use This Code~Generate the Correct Terms for Your Paper}
%\ccsdesc{Do Not Use This Code~Generate the Correct Terms for Your Paper}
%\ccsdesc[100]{Do Not Use This Code~Generate the Correct Terms for Your Paper}



\maketitle

\section{Introduction}
\label{sec:intro}

The emotional impact of music is profound and pervasive, influencing the listener's mood, behavior, and cognitive functions \citep{mozart}. The ability to automatically detect emotions in music has significant implications across various domains, including music recommendation systems, therapeutic applications, multimedia content creation and analysis,  human-computer interaction (HCI), along with other commercial applications. Understanding emotional indicators can enhance user experiences in music streaming services, where personalized recommendations based on emotional preferences can lead to more satisfying and engaging interactions \citep{background}. In therapeutic settings, music is often used as a tool for emotional expression and regulation, making accurate emotion detection crucial for developing effective music-based interventions \citep{therapy}. Additionally, in multimedia content creation and analysis, identifying the emotional tone of music can aid in selecting the right soundtrack to evoke desired emotions in films, advertisements, and other media projects. This precision ensures that the audio complements the visual narrative, amplifying the overall, desired emotional impact \citep{wang2000multimedia}. In HCI, understanding the emotions conveyed by music can enhance the emotional responsiveness and empathy of systems, fostering more immersive and personalized user experiences. By interpreting the emotional intent of music, systems can dynamically adjust to the user's emotional state, offering mood-aligned content, improving engagement, and delivering more supportive experience \cite{2011-MER}. Moreover, in commercial applications, brands can leverage emotion recognition to align their marketing strategies with the consumer's emotional state, enhancing brand resonance and loyalty while fostering more personalized and impactful consumer relationships \cite{music_in_ads}. Artificial intelligence (AI)-based music creation and synthesis further expand creative possibilities by generating original compositions \cite{AI_EMO} and adapting music to specific emotional contexts \cite{AI_EMO_2}, providing a powerful tool for both artistic expression and commercial use. This relies on the ability to detect emotion in music.

Music emotion recognition (MER) is a multidisciplinary field at the intersection of musicology, psychology, and computer science, aiming to identify and classify emotions expressed in music. From a technical standpoint, MER involves the analysis of musical elements such as melody, harmony, rhythm, and lyrics. Traditional methods for MER focus primarily on audio features extracted from the audio signal, but recent advancements emphasize the importance of multimodal approaches, incorporating additional modalities such as lyrics, visuals, user context, and physiological responses \citep{2008_1,paper02, 2017-2, paper13,paper11,paper04}. In recent years, research in MER has witnessed a rapid growth, with the advancement of AI and the increased accessibility of digital music. Multiple surveys have been conducted on MER \citep{ISMIR-2010,2011-MER, 2022-Overall,continuous_survey,icac_survey}, focusing primarily on audio features and traditional machine learning algorithms \cite{ISMIR-2010, 2011-MER, icac_survey} or, more recently, deep learning \cite{2022-Overall}. Systematic evaluations on automated MER as a continuous regression problem in the arousal-valence (AV) plane using audio have also been presented \cite{continuous_survey}. However, past literature lacks an overview of the state-of-the-art in MER using not only audio but also other modalities.

Our paper aims to fill the current literature gap by providing a comprehensive survey on approaches to multimodal MER (MMER). These approaches integrate multiple modalities such as audio, lyrics, video, physiological signals, and others, to achieve more accurate and reliable results. Additionally, being a fast-evolving field, this paper delves into more recent advancements in deep learning-based MMER since 2022, highlighting the latest techniques that leverage multimodal data for more accurate and reliable MER. We start by summarizing the necessary background and preliminaries relevant to understand what MER is about (Section \ref{Background}). Then we discuss the methods and techniques in the process of MMER in terms of the previous literature (Section \ref{methods}). Next, we highlight current trends in MMER, challenges, and future directions (Section \ref{sec:future}). Finally, we summarize the main conclusions and take-home lessons (Section \ref{conclusion}).

\section{Background and Preliminaries} \label{Background}
Before delving deeper into MMER methods and challenges, we first discuss different theories of emotion in music (Section \ref{TEM}), as well as important emotion models (Section \ref{emotion_mods}), the primary modalities used in MMER (Section \ref{modalities}), publicly available datasets for music emotion recognition (Section \ref{datasets}), and metrics for method evaluation (Section \ref{eval}).

\begin{table}[!t]
\centering
\small
%\resizebox{\textwidth}{!}{%
\begin{tabular}{lp{0.6\textwidth}}
\toprule
\textbf{Emotion} & \textbf{Musical Features} \\
\midrule
Happiness & Fast tempo, Small tempo variability,\newline Major mode, Simple and consonant harmony,\newline Medium-high sound level, Small sound level variability,\newline High pitch, Ascending pitch,\newline Perfect 4th and 5th intervals,\newline Staccato articulation \\
\midrule
Sadness & Slow tempo,\newline Minor mode, Dissonance, \newline Low sound level,  Moderate sound level variability,\newline Low pitch, Descending pitch,\newline Small intervals (e.g., minor 2nd), \newline Legato articulation \\
\midrule
Anger & Fast tempo, Small tempo variability,\newline Minor mode, Dissonance, \newline High sound level, Small loudness variability,\newline High pitch, Ascending pitch, \newline Major 7th and augmented 4th intervals,\newline Staccato articulation \\
\midrule
Fear & Fast tempo, Large tempo variability, \newline Minor mode, Dissonance, \newline Low sound level,  Large sound level variability,  \newline High pitch,  Ascending pitch, \newline Staccato articulation \\
\midrule
Tenderness & Slow tempo, \newline  Major mode, \newline Medium-low sound level, Small sound level variability, \newline Low pitch,\newline Legato articulation \\
\midrule
Surprise & Fast tempo, Large tempo variability, \newline Major mode, \newline High sound level, Large sound level variability,\newline High pitch, Sudden dynamic changes, Staccato articulation, \newline Unexpected pauses or rhythmic changes \\
\bottomrule
\end{tabular}%}
\caption{Summary of musical features correlated with discrete emotions. Features for all emotions except ``Surprise'' are adopted from Patrik et al.~\cite{Patrik}.}
\label{tab:emo_mus}
\end{table}

\subsection{Theories of Emotion and Music} \label{TEM}
To design and develop effective computational models for MER, it is crucial to comprehend the connection between emotion and music. Numerous features of music have been identified as indicative of distinct emotions. Basic emotions (Table \ref{tab:emo_mus}) can be conveyed by specific musical features, such as \citep{audio_F}:
\begin{itemize}
    \item Tempo: The speed at which a piece of music is played.
    \item Mode: The scale or tonal framework used in a piece such as major or minor.
    \item Harmony: The simultaneous perception of two or more notes producing a pleasing sound.
    \item Sound Level: The amplitude or intensity at which the musical composition is performed.
    \item Pitch: The position of an individual sound within the entire spectrum of sound.
    \item Interval: The discrete change from one pitch to another (major 3rd, perfect 5th, etc.~\cite{intervals}).
    \item Articulation: The way notes are played in sequence (legato, staccato, etc.~\cite{articulation}).
    \item Rhythm: The pattern of beats and how they are grouped together.
    \item Melody: The sequence of notes that are perceived as a single entity.
    \item Dynamics: The volume of the music, including changes in loudness and softness.
\end{itemize}
However, the same feature can be used in a similar manner to express different emotions. As an example, a fast tempo can be used to express happiness, anger, or fear. Thus, each feature in and of itself is neither necessary nor conclusive. Despite this ambiguity, the greater the number of indicators employed, the more dependable the communication becomes \citep{juslin}.

\subsection{Emotion Models} \label{emotion_mods}
\begin{figure}[!t]
    \centering
    \includegraphics[width=\textwidth]{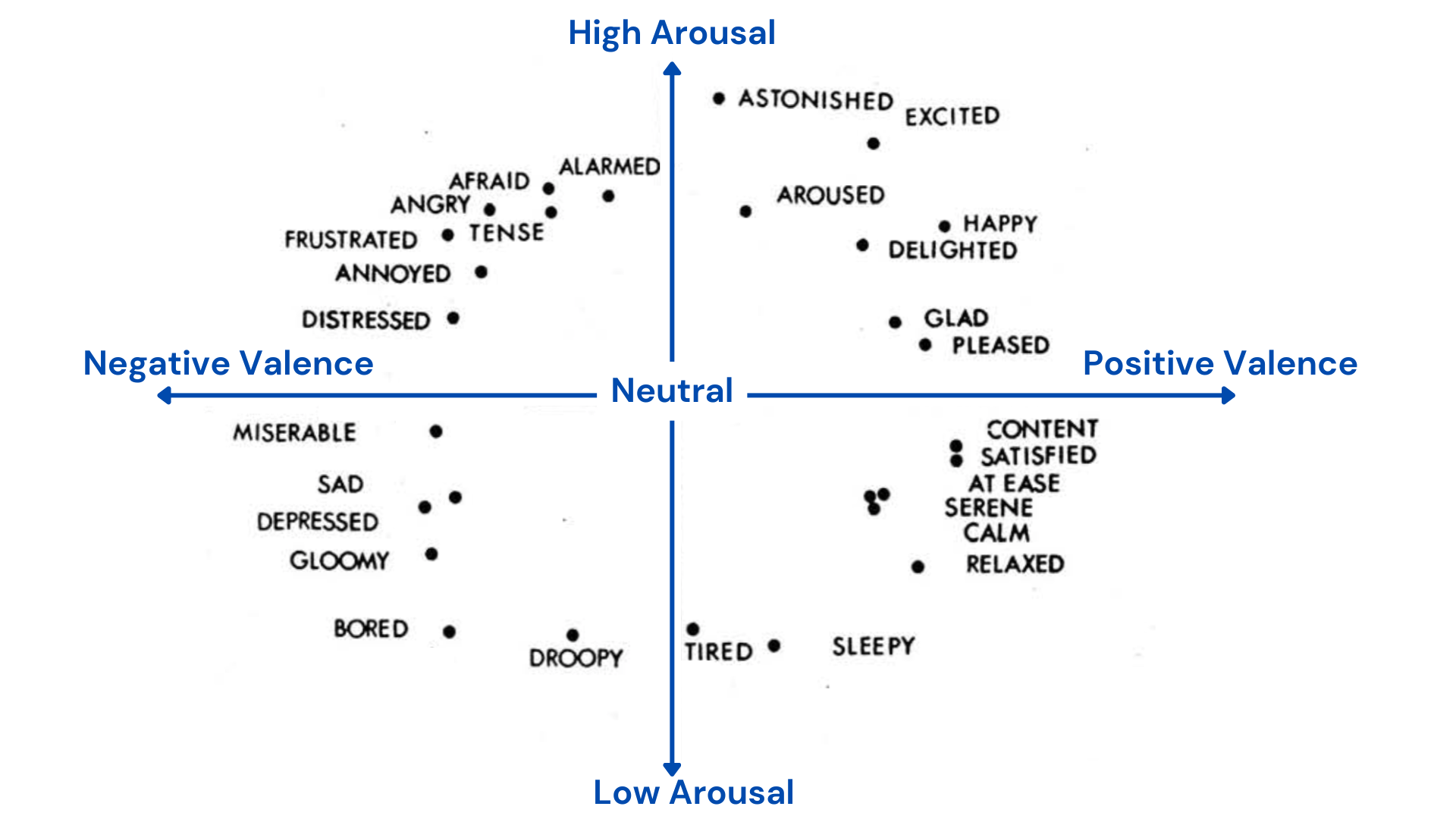}
    \caption{Russell's Circumplex Model. Emotions are characterized by valence (ranging from negative to positive) along the horizontal axis and arousal (ranging from low to high) along the vertical axis.}
    \Description{A circumplex model showing valence on the horizontal axis from negative to positive and arousal on the vertical axis from low to high.}
    \label{fig:russells}
\end{figure}

Understanding and utilizing emotion models is fundamental in MER. Emotion models provide frameworks for categorizing and interpreting the complex and often subjective nature of emotions conveyed through music, facilitating more consistent and comparable results across research and applications. Among the widely recognized models (Table \ref{tab:my_label}), Russell’s Circumplex Model \citep{russell} and Thayer’s Model \citep{thayers} are noteworthy here. Russell’s model maps emotions onto a circular structure defined by valence (expressing feelings from negative to positive) and arousal (expressing the intensity of the emotion from low to high), offering a continuous and dynamic perspective on emotional experiences (Fig.~\ref{fig:russells}). Thayer’s Model, in contrast, introduces the dimensions of energy-stress and calm-tired, focusing on the physiological aspects of emotions.

\begin{table}[!b]
    \centering
    \resizebox{\textwidth}{!}{%
    \begin{tabular}{lccc}
    \toprule
        \textbf{Model} &\textbf{\begin{tabular}[c]{@{}c@{}}Classes /\\Dimensions\end{tabular}} &\textbf{\begin{tabular}[c]{@{}c@{}}Categorical / \\ Dimensional\end{tabular}} & \textbf{Domain} \\ 
        \midrule
         Russell's Circumplex Model \citep{russell} & 2 & Dimensional & General  \\ %\hline
         Thayer's Model \citep{thayers} & 2 & Dimensional &  General \\ %\hline
         Plutchik's Wheel of Emotions \citep{plutchikBook} & 8 & Categorical &  General \\ %\hline
         Ekman's Basic Emotions \citep{ekman1971universals} & 6 & Categorical  & General   \\ %\hline
         Scherer's Component Process Model \citep{schrer} & Varies & Dimensional  &  General \\ %\hline
         Lindquist's Conceptual Act Model \citep{Lindquist} & Varies & Both & General  \\ %\hline
         Positive Activation Negative Activation Model \citep{watson_T} & 2 & Dimensional & General  \\ %\hline
         Geneva Emotion Wheel \citep{GEW} & 20 & Categorical & General   \\ %\hline
         Barrett's Model \citep{barretts} & Varies & Dimensional & General \\ %\hline
         Lazarus's Cognitive-Mediational Theory \citep{Lazarus} & Varies & Both &  General \\ %\hline
         Watson and Tellegen's Circumplex Model of Affect \citep{watson_T} & 2 & Dimensional & General \\ %\hline
         Geneva Emotional Music Scale Model \citep{GEMS} & 9 & Categorical & Music \\ %\hline
         Hevner's Emotional Model \citep{Henver} & 8 & Categorical &  Music \\
         \bottomrule
    \end{tabular}}
    \caption{Emotion models used in MER.}
    \label{tab:my_label}
\end{table}

Other models include Plutchik’s Wheel of Emotions \citep{plutchikBook} and Ekman’s Basic Emotions \citep{ekman1971universals}. Plutchik’s model arranges eight primary emotions in a wheel, highlighting their combinations and interactions, while Ekman’s model identifies six universal emotions: happiness, sadness, fear, disgust, anger, and surprise. These discrete emotion models provide a straightforward approach to categorizing emotional content. More complex models such as Scherer’s Component Process Model (CPM) \citep{schrer} and Lindquist’s Conceptual Act Model \citep{Lindquist} emphasize the dynamic and constructed nature of emotions. CPM focuses on the dynamic changes in emotion components, whereas Lindquist’s model integrates core affect with conceptual knowledge, underscoring the role of cognitive processes in emotional experiences.

Additionally, models such as Positive Activation Negative Activation (PANA)  \citep{watson_T} and Geneva Emotion Wheel (GEW) \citep{GEW} offer nuanced insights into how emotions are constructed and experienced. Barrett’s theory \citep{barretts} suggests that emotions are constructed in the moment by core affect and conceptual knowledge, while the GEW maps complex emotions in a circular manner, reflecting their multidimensional nature. The PANA model distinguishes between positive and negative activation dimensions.

Furthermore, Lazarus' Cognitive-Mediational Theory \citep{Lazarus} highlights the critical role of cognitive appraisal in emotional responses. According to this theory, emotions arise from an individual's evaluation of an event's significance and their ability to cope with it. This appraisal process determines the emotional reaction and subsequent coping strategies, providing a framework for understanding how personal interpretation influences emotional experiences.

Moreover, Watson and Tellegen’s Circumplex Model of Affect \citep{watson_T}, similar to Russell's model, emphasizes positive and negative affects. Each of these models contributes unique perspectives, enhancing our understanding of the intricate relationship between music and emotions.

Two emotion models have been used in the context of music, namely, the Geneva Emotional Music Scale (GEMS) \citep{GEMS} and Hevner's Emotional Model \citep{Henver}. The GEMS model identifies distinct emotional responses such as wonder, transcendence, tenderness, nostalgia, peacefulness, power, joyful activation, tension, and sadness, providing a tailored framework for exploring the rich and nuanced emotional landscape evoked by musical experiences. In contrast, Hevner's model classifies music emotions based on a set of adjectives divided into eight groups. Although this discrete model is less smooth than Thayer’s, it is useful in distinguishing between specific emotional states described in musical terms.

\subsection{Modalities Used in MMER} \label{modalities}
\begin{figure}[!t]
    \centering
    \includegraphics[width=0.8\textwidth]{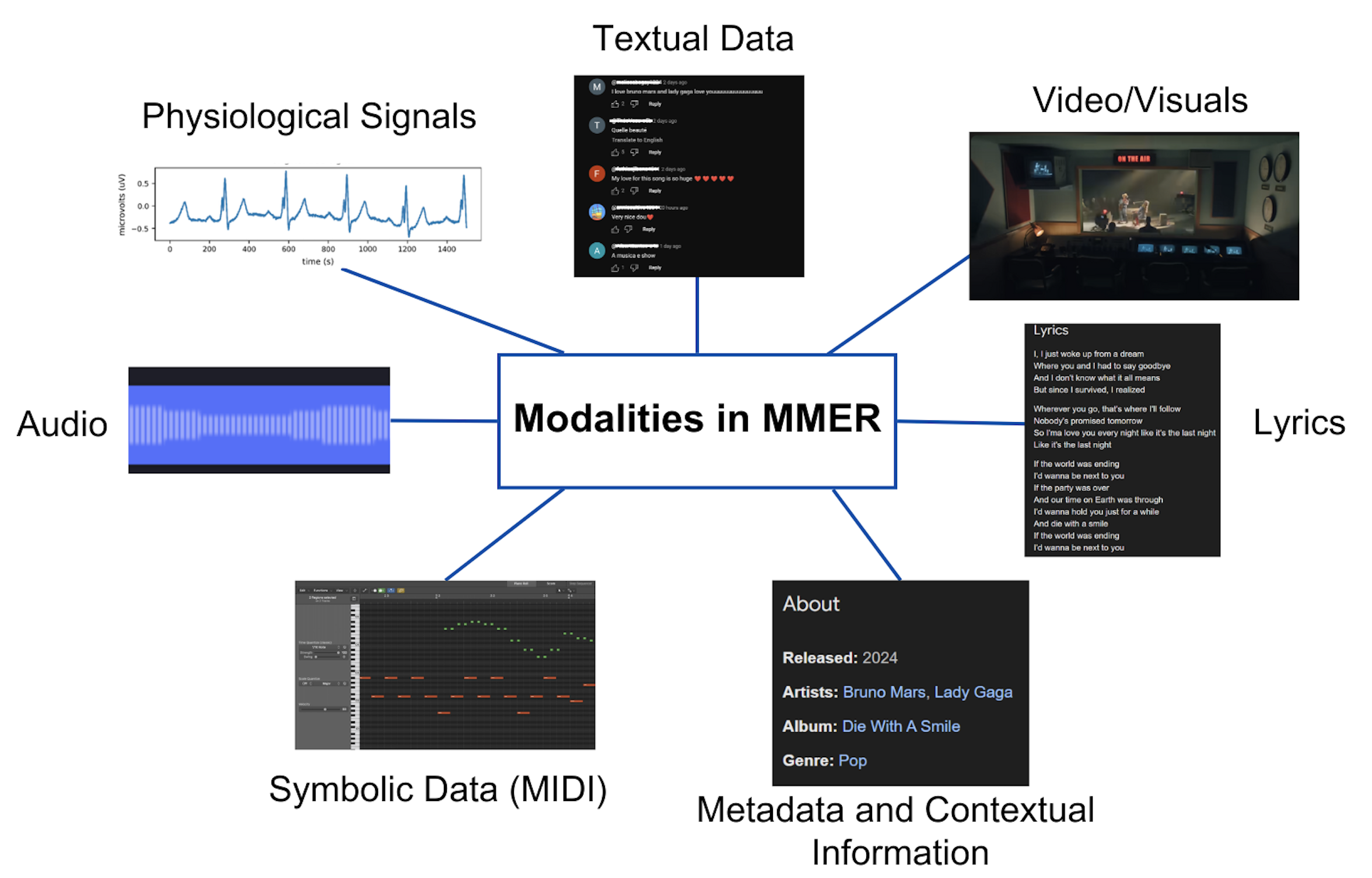}
    \caption{Modalities used in MMER.}
    \Description{A diagram showing different modalities such as audio, lyrics, and visual components that are used in music emotion recognition.}
    \label{fig:mod-class}
\end{figure}

MMER is a multidisciplinary field that uses various modalities to understand and classify the emotional content of music. By combining different sources of information, researchers can develop more accurate and nuanced models. The primary modalities used in MMER are (Fig.~\ref{fig:mod-class}):

\begin{enumerate}
    \item \textbf{Audio:} This refers to the sonic aspect of music, including pitch, loudness, and tempo. It is used to understand and identify the emotions conveyed through the musical piece.

    \item \textbf{Lyrics:} Textual content of songs play a role in conveying underlying emotions. Features such as word choice, themes, and sentiment can be used to identify and interpret the emotions conveyed by the lyrics.

    \item \textbf{Visuals:} Music is often accompanied by videos, live performances, or animations containing visual indicators like color, lighting, facial expressions, and body language, which contribute to the overall emotional experience of the music.

    \item \textbf{Symbolic Data:} MIDI (Musical Instrument Digital Interface) is a mode of music representation that uses discrete events and parameters such as notes, dynamics, and tempo. It enables studying musical structure and patterns to determine emotional content. This symbolic data can be mapped to acoustic features, such as sound intensity, pitch variation, and rhythmic patterns, providing a bridge between the abstract representation of music and its actual auditory characteristics, thereby enhancing the analysis of emotional expression.

    \item \textbf{Physiological Signals:} Emotional responses to music can be measured in the form of changes in skin conductance and heart rate or by electroencephalography (EEG) and electrocardiography (ECG). These measurements provide quantitative data about the physiological effects of music on the listener's qualitative emotional state.

    \item \textbf{Textual Data:} Apart from lyrics, other textual data such as reviews, social media posts, and comments about songs can be used to analyze and understand the emotional impact of music. They capture subjective emotional responses and sentiments from listeners.

    \item \textbf{Metadata and Contextual Data:} Further information about music, such as genre, artist, release year, and cultural context, is often provided in the metadata of digital songs. The background and contextual understanding to which this information contributes enhances the emotional interpretation of a song.

    \end{enumerate}

\begin{table}[!t]
\centering
\resizebox{\textwidth}{!}{%
\begin{tabular}{lccclc}
\toprule
\textbf{Dataset} & \textbf{Year} & \textbf{Size} & \textbf{Annotation} & \textbf{Modalities} & \textbf{Labeling} \\ 
\midrule
% \hline
CAL500 \citep{CAL500} & 2008  & 500 & SLEA  & Audio  & Categorical                                                            \\ 
% \hline
Soundtracks \citep{soundtracks}  & 2011                                                           & 470                                                             & SLEA             & Audio                                                                                                                & Both \\ 
% \hline
DEAP    \citep{deap}     & 2012                                                           & 120                                                             & SLEA             & Video, EEG, Physiological                                    & Dimensional                                                                 \\ 
% \hline
MediaEval Emotion in Music \citep{MediaEval} & 2013 & 744                                                            & Both               & Audio                                                                                                                & Dimensional                                                                 \\ 
% \hline
MIREX-like \citep{mirex-like} & 2013                                                       & 193                                                             & SLEA             & Audio, Lyrics, MIDI                                                    & Categorical                                                                 \\ 
% \hline

CAL500exp \citep{CAL500exp}   & 2014                                                            & 3,223                                                            & CEA               & Audio                                                                                                                & Categorical                                                                 \\ 
% \hline

AMG1608 \citep{AMG1608}  &                       2015                                       & 1,608                                                            & SLEA             & Audio                                                                                                                & Dimensional                                                                 \\ 
% \hline
NJU   \citep{2015_1}    & 2015                                                              & 777                                                             & SLEA             & Audio, Lyrics                                                            & Categorical                                                                 \\ 
% \hline
Emotify     \citep{emotify}        &                   2016                                 & 400                                                             & SLEA             & Audio                                                                                                                & Categorical                                                                 \\ 
% \hline

Audioset  \citep{audioset}       &2017                                                       & 16,955                                                           & SLEA             & Video                                                                                                                & Categorical                                                                 \\ 
% \hline
DEAM \citep{DEAM}           & 2017                                                        & 1,802                                                            & Both               & Audio, Metadata                                                           & \multicolumn{1}{c}{Dimensional}                                            \\ 
% \hline

4Q Dataset  \citep{4Q}  & 2018                                                                & 900                                                             & SLEA             & Audio                                                                                                                & Categorical                                                                 \\ 
% \hline
PMEmo   \citep{PMEmo} & 2018                                                               & 794                                                             & Both               & Audio, Lyrics, Comments, Physiological & Dimensional \\ 
% \hline
The MTG-Jamendo \citep{mtg} &  2019 &  55,525                                                             & SLEA           & Audio                                                     & Categorical \\ 
% \hline
Music4all \citep{music4all}  & 2020                                                           & 109,269                                                             & Both          & Audio                                                    & Dimensional                                                                 \\ 
% \hline
EMOPIA \citep{emopia}  & 2021                                                           & 1,087                                                             & SLEA           & MIDI, Metadata & Categorical                                                                \\ 
% \hline

MuSe \citep{MuSe}    & 2021                                                         & 90,001                                                             & SLEA           & Audio                                                  & Dimensional                                                                 \\ 
% \hline
HKU956 \citep{HKU956}   &2022 & 956                                                             & SLEA           & Audio, Physiological signals                                                     & Dimensional                                                                 \\ 
% \hline
MusAV \citep{musAV}   & 2022 & 2,092                                                             & SLEA           & Audio                                                    & Dimensional                                                                 \\ 
% \hline

MuVi  \citep{muVi}   &2022                                                          & 81                                                             & CEA           & Audio, Video (Muted) & Dimensional                                                                 \\ 
% \hline
TROMPA-MER \citep{Tompa}    & 2022                                                         & 1,161                                                             & SLEA           & Audio                                                 & Categorical                                                                 \\ 
% \hline
emoMV   \citep{emoMV}    & 2023                                                            & 5,986                                                            & SLEA             & Audio, Video                                                         & Categorical                                                                 \\ 
% \hline

MERP \citep{MERP}     & 2023                                                               & 54                                                              & CEA               & Audio, Metadata                                                         & \multicolumn{1}{c}{Dimensional}                                            \\ 
% \hline
EMMA \citep{EMMA}         & 2024                                                    & 364                                                             & SLEA           & Audio, Video                                                    & Categorical \\ 
% \hline
SiTunes \citep{siTunes}        & 2024                                                     & 300                                                             & SLEA           & Audio, Physiological, Weather & Dimensional \\ 
% \hline
MERGE \citep{merge}        & 2024                                                     & 2,000                                                             & SLEA           & \begin{tabular}[c]{@{}l@{}}Audio, Lyrics\end{tabular}                                                      & Dimensional \\ 
% \hline   
\bottomrule
\end{tabular}}
\caption{Publicly available MER datasets. Size indicates the number of songs or fragments in the dataset. Annotation includes SLEA (song-level emotion annotation), CEA (continuous emotion annotation), or both. Labeling includes categorical (emotions classified into distinct categories such as happy or sad), dimensional (emotions represented on a continuous scale such as valence or arousal), or both.}
\label{tab:datasets}
\end{table}

\subsection{Music Emotion Recognition Datasets} \label{datasets}
Many datasets are available for MER research (Table \ref{tab:datasets}). However, most of them are unimodal, typically containing only audio. Broadly speaking, existing datasets can be divided into two categories: song-level emotion annotation (SLEA) and continuous emotion annotation (CEA). The annotation labels used in each case can be categorical (distinct emotions) or dimensional (numerical values over a number of emotion dimensions such as arousal and valence \cite{russell}). The latter is particularly useful for capturing the nuanced and dynamic nature of musical emotions.

\subsubsection{Song-Level Emotion Annotation (SLEA)}
Most datasets use song-level annotation, where a single emotion is assigned to a complete song. These datasets typically contain annotations that label each song with the predominant emotion. The assumption that a song contains a single emotion may not always be true but may work well as a simplification when using MER in a downstream task such as music information retrieval or recommendation.

\subsubsection{Continuous Emotion Annotation (CEA)}
In contrast to song-level annotation, continuous annotation provides time-varying labels that capture the dynamic nature of emotions throughout a song. The time intervals for these annotations are typically determined based on the granularity required for accurate emotion tracking, often ranging from 1 to 5 seconds. Shorter intervals provide a finer resolution of emotional changes but require more detailed analysis, while longer intervals may capture broader emotional trends but with less temporal precision. The choice of interval depends on the specific objectives of the study and the nature of the music content being analyzed. These datasets are crucial for studies on music emotion variation detection (MEVD).

\subsection{Evaluation Metrics} \label{eval}
Evaluation metrics are essential for assessing the performance and robustness of MMER methods. MMER may be modeled as a classification or a regression task. For the classification formulation of MMER, Accuracy (A) is a commonly used metric representing the probability of correct classification within a dataset, and provides a basic measure of overall correctness. It is calculated as the ratio of correctly predicted outcomes to the total number of predictions made. Precision (P) and recall (R) are useful for evaluating the reliability and completeness of positive predictions, respectively. Precision measures the ratio of correctly predicted positive observations to the total predicted positives, while recall measures the ratio of correctly predicted positives to all actual positives. The F1 score, being the harmonic mean of precision and recall, is particularly useful for handling imbalanced datasets, offering a balanced view of the model's performance \citep{metrics1,metrics2}.

For regression tasks in MMER, metrics such as the mean absolute error (MAE), the coefficient of determination (R$^2$), and the root mean squared error (RMSE) are frequently used. MAE estimates the average magnitude of prediction errors, allowing a direct interpretation of the average error. R$^2$ evaluates the degree to which a given regression model accurately represents the sample data. RMSE, on the other hand, prioritizes greater errors, making it more vulnerable to outliers. The concordance correlation coefficient (CCC) is also used to assess the agreement between predicted and actual values by integrating precision and accuracy into a single metric. Additionally, area under the receiver operating characteristic curve (AUROC) is employed to evaluate the classifier’s ability to distinguish between classes at various threshold settings, providing a comprehensive measure of separability. In ranking tasks, metrics such as hits score on top k (hits@k) and mean average precision on top k (map@k) are utilized to evaluate the accuracy and order of a model’s top predictions, respectively. These metrics collectively provide a robust framework for evaluating the diverse aspects of MMER performance \citep{2022-Overall,metrics_reg}.

Furthermore, public competitions and challenges provide researchers in the field with benchmarks to evaluate and compare methods. These challenges utilize standardized datasets, consistent evaluation protocols and metrics, and allow competitors to rank their methods against the state of the art. Currently, there are no established benchmarks developed specifically for MMER. The only available benchmarks are for audio-based MER. Some of these include the ``Emotion and Themes in Music'' task in MediaEval\footnote{\url{https://multimediaeval.github.io/editions/2021/tasks/music/} (Accessed 15 November 2024).} (most recent: 2021) and the ``Audio K-POP Mood Classification'' task in MIREX\footnote{\url{https://www.music-ir.org/mirex/wiki/2020:Audio_K-POP_Mood_Classification} (Accessed 15 November 2024).} (most recent: 2020). However, both are based on audio classification only.

Selecting the appropriate evaluation metrics is very important for effective performance assessment and improvement. The chosen metrics should align with the task-specific characteristics and the nature of the dataset. To obtain a comprehensive understanding of the strengths and weaknesses of their models, researchers can utilize a combination of the discussed metrics.

\section{Methods and Techniques} \label{methods}

\begin{figure*}[!t]
  \centering
   \includegraphics[width=1\textwidth]{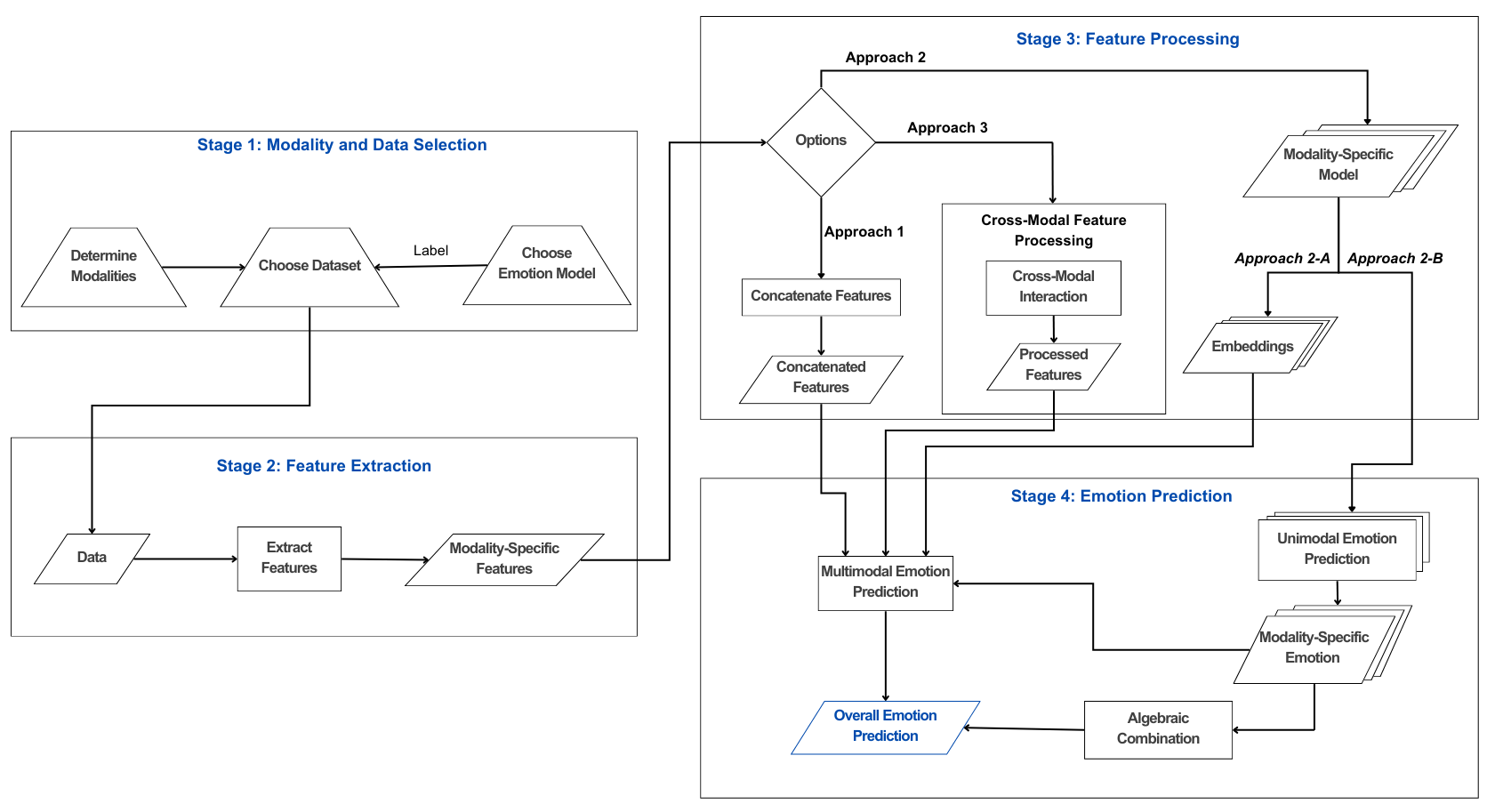}
  % \includesvg[inkscapelatex=false, width=\textwidth]{Images/SVG/MO_FInal_new.svg}
  \caption{Framework summarizing past and current MMER methods.}
  \Description{A flowchart-like framework summarizing various approaches and modalities in multimodal music emotion recognition (MER).}
  \label{fig:framework}
\end{figure*}

MMER research can be summarized using a four-stage framework (Fig.~\ref{fig:framework}), proceeding from modality and data selection (Stage 1), to feature extraction (Stage 2) and feature processing (Stage 3), and, finally, emotion prediction (Stage 4). While Stages 1 and 2 exhibit significant similarities across all published works, Stage 3 typically uses one of three distinct approaches, while Stage 4 commonly involves one of two distinct approaches. Having already introduced the various aspects of Stage 1 (especially in Sections \ref{emotion_mods}--\ref{datasets}), we now proceed to discuss methods and techniques involved in Stage 2 (Section \ref{sec:feature_extraction}), Stage 3 (Section \ref{sec:feature_processing}), and Stage 4 (Section \ref{sec:feature_fusion}).

\subsection{Feature Extraction} \label{sec:feature_extraction}
Feature extraction is a crucial step in MMER, as it transforms data from different modalities into representations of relevant information about a given piece of music that can be used as a basis for further processing and ultimate prediction. In this section we present a detailed overview of feature extraction in MMER, including the data formats, methods and techniques used to obtain audio features, lyric features, visual features, symbolic features, physiological features, textual (user-generated content) features, and metadata/contextual features.

\subsubsection{Audio Features}
\begin{figure}[!t]
    \centering
    \includegraphics[width=0.35\linewidth]{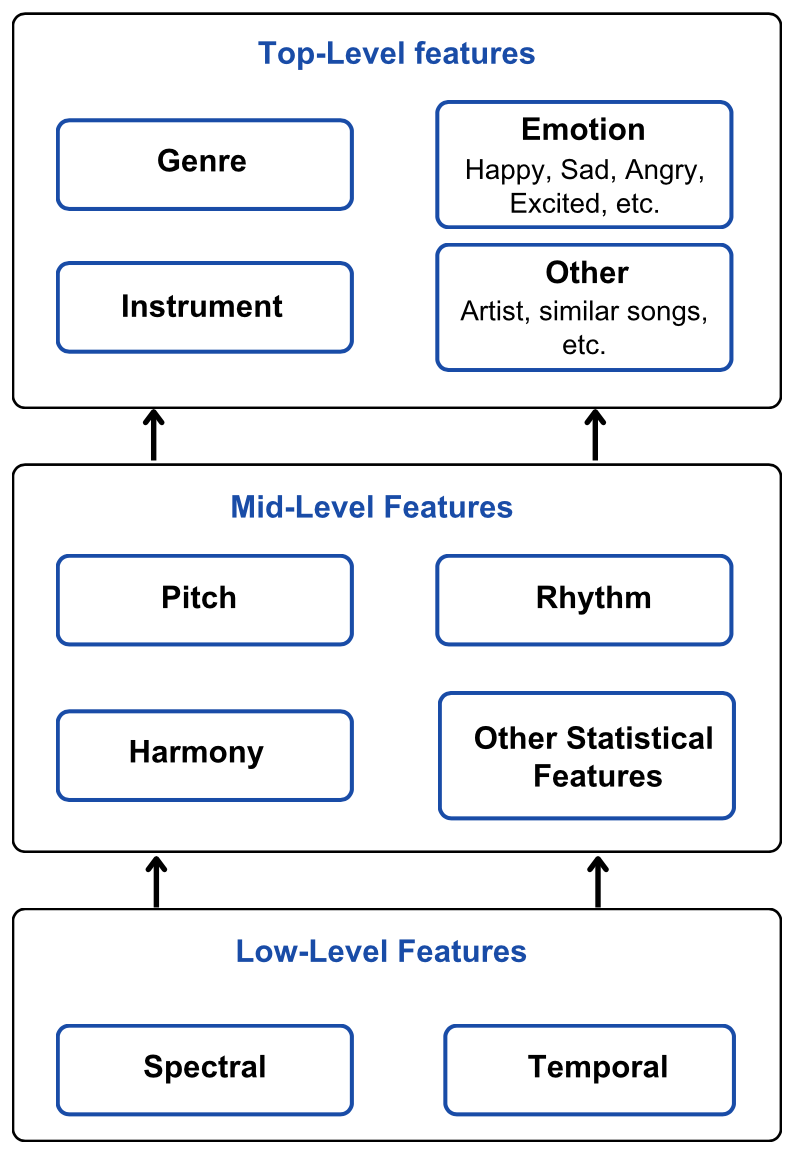}
    \caption{Categorization of audio features.}
    \Description{A chart or diagram categorizing different audio features used in music analysis, such as timbre, rhythm, and melody.}
    \label{fig:audio_features_cat}
\end{figure}

These features are the most widely studied in MER as well as MMER. The key audio elements conveying or expressing emotion in music are tempo, rhythm, mode, melody, and dynamics. Fu et al.~\cite{Zhouyu} divide audio features into three categories: low-level, mid-level, and top-level (track-level) features/labels (Fig.~\ref{fig:audio_features_cat}). Semantic labels, at the top-level, offer insights into the ways in which people perceive and interpret music like style, genre, emotion, etc. As obtaining these levels is difficult from lower-level features, it makes the top-level features abstract \cite{features_N_1}. Therefore, in this survey, we primarily focus on low- and mid-level features, as these serve as the foundation for extracting top-level features like emotion. Top-level features and fusion are included in the discussion of emotion prediction (Section \ref{sec:feature_fusion}), aligning with the ultimate objective of understanding emotional responses to music.

\paragraph{Low-Level Features} These can be broadly categorized as spectral features (SFs) and temporal features (TFs). The spectral characteristics of music include the timbre and tonal properties and are captured by SFs present in a relatively short time interval \citep{Jeong}. SFs are obtained using signal processing techniques such as Fourier transformation, spectral/cepstral analysis, etc. Furthermore, the SFs commonly used in MER consist of two types: spectrograms and low-level descriptors (LLDs). The short-time Fourier transform (STFT) is used to generate spectrograms, and the input for models using these is in the form of a two-dimensional (2D) map consisting of a sequence of short-time spectrograms.  LLDs calculated using single-frame audio are further processed to obtain high-level statistical functions (HSFs) \citep{paper07}. Examples of LLDs are the mel-frequency cepstrum coefficient (MFCC) \citep{logan}, zero crossing rate (ZCR), spectral centroid, spread, roll-off, and flux, and chroma features \citep{audio_Feat}. Examples of HSFs are the maximum, mean, and variance. In recent literature, the most commonly used timbre features are MFCCs, ZCR, and chroma features, while the centroid is the most widely used spectral feature (Table \ref{tab:audio_features}). However, some authors have used unique features, such as joined filter banks (an intermediate product between mel-spectrograms and MFCCs) and 60 handcrafted features to generate input for the classification model \cite{paper18}.

TFs, on the other hand, capture the time-related aspects of audio signals, such as rhythm, tempo, and energy dynamics that are essential for expressing emotions in music. These features capture aspects such as beat consistency, tempo variations, and loudness fluctuations. Examples of TFs include short-time energy (STE), tempo, and beat strength, which characterize the rhythmic and dynamic elements of a musical piece. Unlike SFs, TFs focus on how these audio characteristics evolve over time. For example, the progression of energy or the consistency of rhythmic patterns can significantly affect how arousal or emotional intensity is perceived \citep{2011-MER}.

The integration of TFs with SFs enhances the representation of musical content by combining instantaneous timbral information with temporal progression, leading to more effective emotion recognition systems. Whether considering SFs or TFs, advanced methods such as convolutional neural networks (CNNs) and recurrent neural network (RNNs) automatically learn hierarchical feature representations, capturing complex patterns and temporal dependencies at multiple scales that are indicative of emotions \citep{choi}.

\paragraph{Mid-Level Features}
In MMER, the use of mid-level features is not as common as in other areas such as cover-song detection. The three most used mid-level features are rhythm, pitch, and harmony. Rhythm is described using two important indicators: beat and tempo (beats-per-minute). The tempo can correspond with other features such as pitch, which is also a mid-level feature, in order to recognize an emotion (see Table \ref{tab:emo_mus}).

\subsubsection{Lyric Features} The lyrics of a song play a crucial part in expressing its emotional charge. Most research uses natural language processing (NLP) models as classifiers or regressors. Therefore, the words of the lyrics need to be converted into specific features to be fed as input to these models. Pyrovolakis et al.~\cite{paper01} used several methodologies such as bag of words (BoW), term frequency-inverse document frequency (TF-IDF), Word2Vec, GloVe, and bidirectional encoder representations from transformers (BERT) embeddings. While BoW is one of the most widely used features in MER, part-of-speech (PoS) tags are also used as features that attribute the grammatical class to each word. However, according to Malheiro et al.~\cite{malheiro}, BoW and PoS are not sufficient for MER. Therefore, they proposed three new features: presence of slang or colloquial words, structural analysis features, and semantic features. The sequence of words (unigram, bigram, and trigram) is another type of lyric features used in MER research \citep{unigram}. Chen and Li \cite{paper07} used two main features: word embeddings and word frequency vectors using Word2Vec and chi-squared test feature extraction respectively. Similarly, Zhang and Tian \cite{paper10} used chi-squared test feature extraction along with TF-IDF to extract the lyric features to train their models. To enhance the previously mentioned techniques, particularly for handling multi-emotion classification problems, Edmonds and Sedoc \cite{edmonds} utilized the NRC Hashtag Emotion Lexion \cite{Mohammad} to convert lyrical data into feature vectors with a length of~9. Wang et al.~\cite{41_survey} incorporated rhyme as a feature using a rhyme system they created for their study on emotion detection in Chinese songs. 

\begin{table}[!t]
\centering
\small
\begin{tabular}{lccc}
\toprule
\multicolumn{1}{l}{{\textbf{Reference}}} & \multicolumn{2}{c}{\textbf{Feature}} & \multicolumn{1}{c}{{\textbf{Year}}} \\ \cmidrule{2-3}
\multicolumn{1}{c}{} & \multicolumn{1}{c}{\textbf{Timbre}} & \multicolumn{1}{c}{\textbf{Spectral}} & \multicolumn{1}{c}{} \\ 
\midrule
Chen and Li \cite{paper07} & \multicolumn{1}{c}{\begin{tabular}[c]{@{}c@{}}MFCC, ZCR, Chroma\end{tabular}} & \begin{tabular}[c]{@{}l@{}}Spectral Centroid, Spectral Spread,\\ Spectral Roll-Off, Spectral Flux\end{tabular} & 2020 \\ 
\midrule
Liu and Tan \cite{paper08} & \multicolumn{1}{c}{MFCC} & Spectrum Centroid & 2020 \\ 
\midrule
Pandeya et al. \cite{paper02} & \multicolumn{1}{c}{\begin{tabular}[c]{@{}c@{}}MFCC, ZCR\end{tabular}} & \multicolumn{1}{c}{-}  & 2021 \\ 
\midrule
Chen \cite{paper12} & \multicolumn{1}{c}{MFCC} & \multicolumn{1}{c}{-} & 2022 \\ 
\midrule
Liu et al. \cite{paper20} & \multicolumn{1}{c}{\begin{tabular}[c]{@{}c@{}}MFCC, ZCR, Chroma\end{tabular}} & \begin{tabular}[c]{@{}l@{}}Spectral Centroid, Spectral Roll-Off,\\ Spectral Flatness, Spectral Contrast\end{tabular} & 2022 \\ 
\midrule
Pyrovolakis et al. \cite{paper01} & \multicolumn{1}{c}{\begin{tabular}[c]{@{}c@{}}MFCC, Chroma\end{tabular}} & Spectral Contrast & 2022 \\ 
\midrule
Zhang and Tian \cite{paper10} & MFCC, ZCR, Chroma  & Spectral Centroid, Spectral Bandwidth & 2022 \\ 
\midrule
Zhang et al. \cite{paper18} & \multicolumn{1}{c}{Filter Bank} & \multicolumn{1}{c}{-} & 2022 \\ 
\midrule
Zhao and Yoshii \cite{paper11} & \multicolumn{1}{c}{MFCC} & \multicolumn{1}{c}{-} & 2023 \\ 
\midrule
Wang et al. \cite{paper03} & \multicolumn{1}{c}{MFCC} & Spectral Centroid & 2024 \\ 
\bottomrule
\end{tabular}
\caption{Summary of audio features used in recent literature (2020-2024).}
\label{tab:audio_features}
\end{table}

\subsubsection{Visual Features} In recent developments within MMER, researchers have begun incorporating visuals, particularly music videos, into their analysis. This addition allows for a richer and more nuanced understanding of the emotional content of music by combining auditory and visual cues, thereby enhancing the accuracy of emotion recognition systems. Visual features can be divided into two main categories based on the input type: static and dynamic. Appearance-based feature extraction approaches are commonly used for static images. Gabor wavelet representation and local binary patterns (LBP) are two of these methods. Recently, researchers have used CNNs to extract facial features from static images \citep{cnn_v_1,cnn_v_2,cnn_v_3}. With regard to dynamic features, Pandeya et al.~\cite{paper02} proposed a method for analyzing video segments that contain only faces. They extracted these using the cascade classifier feature of OpenCV\footnote{\url{https://opencv.org/} (Accessed 15 November 2024).}. Facial expression is a visual feature used in the majority of MER literature. Chen \cite{paper12} used the 68 face feature points obtained using the Dlib\footnote{\url{https://pypi.org/project/dlib/} (Accessed 15 November 2024).} Python library to extract these facial expressions. A common strategy to capture the dynamics of movement within a video is to employ optical flow and motion analysis \citep{optic_flow_motion}. To infer emotions more accurately, researchers have used context-aware networks that consider the scene's broader context, including the environment and character interactions \citep{paper02}. Additionally, slow-fast networks, which process video data at multiple temporal resolutions, capture both detailed spatial information and rapid temporal changes \citep{paper20}.

\subsubsection{Symbolic Features} These features are extracted from symbolic music scores such as MIDI. Most symbolic music scores used in MER are represented in MIDI, which is an informative resource that provides note information (pitch, duration, etc.), timing information (tempo and time signature), and other relevant details. However, in the case of multimodal or unimodal MER, MIDI files are subjected to additional processing in order to extract symbolic domain features such as melody, rhythm, and other dynamic information. Zhao and Yoshii \cite{paper11} utilized a RNN, specifically a bidirectional gated recurrent unit (BiGRU), to extract symbolic domain features from MIDI in order to pass them to the emotion classifier. Others, such as Thammasan \cite{2017-2}, employed the MIRToolbox developed by Laurier et al.~\cite{MIRtoolbox} to extract musical features from MIDI files.

\subsubsection{Physiological Features} These can be derived from several sources, including EEG, ECG, heart rate variability (HRV), galvanic skin response (GSR), electromyography (EMG), and respiratory rate (RSP) \citep{psycho} in both multimodal and unimodal MER.

EEG, in particular, is widely used in the field of MER \citep{EEG_survey} as it captures the brain's electrical activity and provides insights into the emotional responses elicited by music. To extract emotion-related features, it is important to preprocess high-dimensional EEG signals, which may include a significant number of irrelevant features. In literature, the most popular EEG features for emotion detection are fractal dimension (FD), power spectral density (PSD), Higuchi fractal dimension (HFD), multiscale fractal dimension (MFD), spectrograms, and the wavelet transform (WT) \citep{EEG_survey}. Some of the methods and tools used to preprocess EEG and obtain these features are \cite{EEG_survey} the Higuchi algorithm \citep{FD} for FD, the STFT, WT, and the Hilbert-Huang Transform (HHT) for PSD, and PyEEG \citep{pyegg} for HFD. Wavelet domain features can be extracted using the STFT, high-order crossing analysis (HOC), and hybrid adaptive filtering (HAF). To remove eye movements, facial muscle movements, heartbeats, and other distortions, independent component analysis (ICA) and blind source separation (BSS) can be used \citep{EEG_survey}.

Similar to EEG, ECG signals have also been used for MER. Hsu et al.~\cite{ECG} developed the sequential forward floating selection-kernel-based class separability (SFFS-KBCS) feature selection algorithm and also utilized generalized discriminant analysis (GDA) in order to select significant ECG features for emotion detection. Furthermore, Naji et al.~\cite{ECG2} calculated several features using the RR (time intervals between consecutive R peaks) time frames extracted from the ECG signal. Some of these features are statistical features (e.g.\ mean and standard deviation), nonlinear features (e.g.\ sample entropy), and triangular phase space mapping.

Hu et al.~\cite{psy_1} defined the relationship between music-induced emotions and physiological signals. They collected electrodermal activity (EDA), blood volume pulse (BVP), inter-beat interval (IBI), heart rate (HR), and skin temperature (TEMP) data and obtained a set of features that can be used in MER (see Table \ref{tab:psyc_sum} for a summary of these and other physiological features).
    
\begin{table}[!t]
\resizebox{\textwidth}{!}{%
\begin{tabular}{ll}

% \hline
\toprule
\textbf{Category} & \textbf{Features} \\ 
\midrule
Statistical Features & Mean, Median, Standard Deviation, Range \\ 
\midrule
Time Domain Features & Mean Absolute Value of 1st/2nd Differences of Raw/Normalized Signals\\
& Fractal Dimension (FD) of EEG \\ 
\midrule
Frequency Domain Features & Low and High Frequency (LF, HF), Ratio LF/HF \\ 
\midrule
Signal-Specific Features & Fractal Dimension (FD), Power Spectral Density (PSD) \\
 & Higuchi Fractal Dimension (HFD), Multiscale Fractal Dimension (MFD) \\
 & Galvanic Skin Response (GSR), Heart Rate Variability (HRV) \\
\bottomrule
\end{tabular}}
\caption{Categorization of physiological features.}
\label{tab:psyc_sum}
\end{table}

\subsubsection{Textual Features} Beyond lyrics, other forms of textual data, such as user reviews, social media comments, and textual descriptions of music, have been employed to gauge emotional responses to music. These texts often capture subjective experiences and sentiments of listeners, which can be analyzed using NLP techniques to infer the emotional impact of music. These features are also extracted similarly to other textual data discussed above \citep{paper25}.

\subsubsection{Metadata and Contextual Features} Methods employing metadata and contextual information such as song title and artist name as a modality for their final prediction, typically use pretrained models such as BERT to extract the features \cite{paper13}. 

\subsection{Feature Processing}\label{sec:feature_processing}
After features have been extracted (Stage 2) from the raw input data (Stage 1), they typically need to be further processed (Stage 3) before they can be used effectively for final emotion prediction (Stage 4). There are three main feature processing approaches that researchers have used up to now (Fig.~\ref{fig:framework}), which we summarize in this section.

\subsubsection{Approach 1: Feature Concatenation}
The most straightforward approach is to concatenate the modality-specific features into a unified feature set. Previous research has explored various techniques to accomplish this. Some studies employ a simple vector representation to merge all features within the same space \citep{2008_1}, while others adopt more advanced methods, such as utilizing a bimodal deep Boltzmann machine (DBM) for feature fusion \cite{Huang}. This approach is related to what is generally called early fusion in the deep learning literature \citep{Ramachandram-2017, Hussain-2024}.

\subsubsection{Approach 2: Modality-Specific Feature Processing} \label{emotion_reg}
An alternative approach is to train modality-specific models and then combine their outputs. There are two variants of this approach, which we refer to as Approach 2-A and Approach 2-B (Fig.~\ref{fig:framework}). In Approach 2-A, each modality-specific model produces an embedding, which is used as input to the final multimodal emotion prediction model. This is known as intermediate fusion \citep{Ramachandram-2017, Hussain-2024}. Most works using this approach train unimodal models, taking the extracted features as input, and then the last few layers of the models are removed to obtain the embeddings \citep{Delbouys}. Alternatively, the unimodal models can be used directly as in Approach 2-B, providing unimodal predictions that can be aggregated to produce a final multimodal prediction. This situation corresponds to late fusion \citep{Ramachandram-2017, Hussain-2024}.

\subsubsection{Approach 3: Cross-Modal Feature Processing} \label{cross_mod}
The most sophisticated approach is to perform cross-modal feature processing. This concept is increasingly adopted by the field of MER and is especially gaining popularity in MMER. Generally speaking, cross-modal processing combines inputs from different modalities to create a single output. This output is then used as the input for the final classifier, regressor, or fusion model. Unlike multimodal processing, which focuses on integrating multiple types of data to improve overall model performance, cross-modal processing emphasizes how interactions between different modalities can enhance or influence the processing of each data type, creating a more cohesive and contextually aware representation. This takes the concept of intermediate fusion to a higher level.

Recent works \citep{paper03, paper04,paper13} have used cross-modal processing to assess the interaction between audio and lyrics. The input consists of pairs of lyric and audio fragments, where each input is typically a single sentence of lyrics and the corresponding audio, and the output is an emotion feature vector. A key element of cross-modal processing in these works is the use of an emotion long-short-term memory (E-LSTM) cell, an enhancement of the traditional LSTM, which processes the next lyric-audio pair along with the emotion vector of the past pair to maintain a consistent emotional state with respect to the previous interaction and to avoid the emotion independence between the two channels. The historical emotion vector is designed to preserve intense emotional information, while the current emotion vector updates relatively weaker emotions. This combined emotion vector is iteratively updated with the historical emotion vector, which helps decide which emotional levels in the song should be maintained and which past emotions need to be revised (see \cite{paper03} for detailed mathematical formulations of the operations in this process).

\subsection{Emotion Prediction}\label{sec:feature_fusion}
\begin{figure}[!t]
    \centering
    \includegraphics[width=\textwidth]{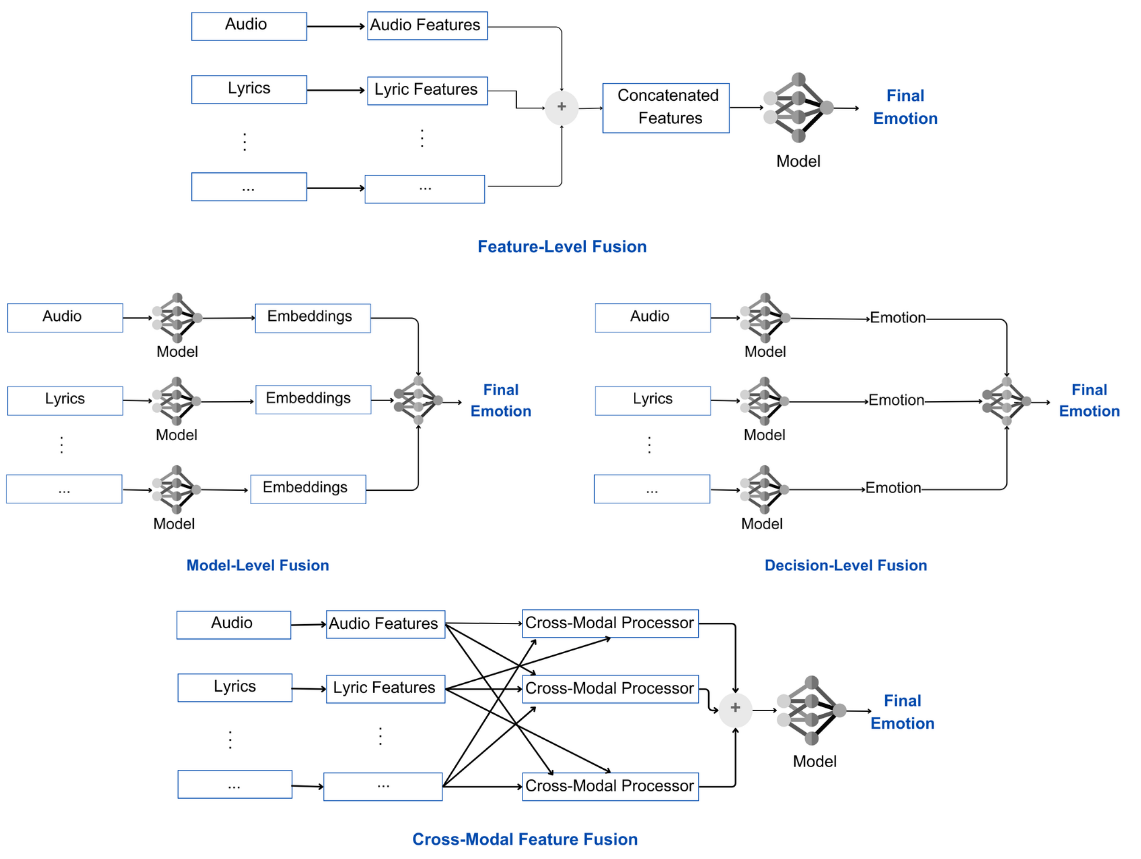}
    \caption{Comparison of fusion methods in music emotion prediction.}
    \Description{A visual comparison of various fusion methods used in data analysis, highlighting their differences and applications.}
    \label{fig:decisions}
\end{figure}

Given the processed features (Stage 3), the final task (Stage 4) is to predict the emotion they convey. The methods used to accomplish this task depend on the form in which the processed features are presented. As alluded to in the previous section, this corresponds to different fusion strategies (Fig.~\ref{fig:decisions}), which we elaborate on here as we discuss their usage in ultimate emotion prediction (see Table \ref{tab:full_summary} for an overview of state-of-the-art MMER methods and performance).

\subsubsection{Emotion Prediction by Feature-Level Fusion}
Feature-level fusion, also known as early fusion, involves combining all extracted features from different modalities into a single high-dimensional feature vector (Approach 1). This unified feature vector is then used to train a single classifier such as support vector machines (SVMs) \cite{2008_1, 2015_1}, auto encoders \cite{2019_01}, or CNNs \cite{paper11, paper06} for emotion prediction. A common approach to feature-level fusion is feature concatenation, where features from modalities such as audio and lyrics are merged into one vector and fed into a classification model. As this creates a multimodal feature space, challenges arise in integrating heterogeneous features from different modalities into a cohesive representation \citep{2008_2}. Researchers have explored various techniques to address these challenges, including dimensionality reduction and normalization, which aim to better align the feature vectors of varying modalities. Dimensionality reduction methods such as principal component analysis (PCA) aim to reduce the number of features while retaining the most important variance in the data, and normalization methods such as z-score normalization and min-max scaling are often applied to ensure that features from different modalities are on a similar scale, which improves the performance of classification algorithms \cite{ISMIR-2010}. Despite these efforts, direct concatenation can sometimes fail to preserve critical emotional information due to the inherent differences between the representations of modal features \citep{paper07}.

\subsubsection{Emotion Prediction by Decision-Level Fusion}
Decision-level fusion, also known as late fusion, combines the outputs of unimodal classifiers or regressors (Approach 2-B). These outputs can be aggregated using algebraic combination rules such as ``Min,'' ``Max,'' or ``Sum'' \citep{paper01, paper07, huertas, paper02, paper08, paper10}. Methods in this category can be further divided into linear probability fusion (LPF) and stacking ensemble learning (SEL) \citep{paper07}. While LPF is widely used in general machine learning applications \citep{LPF}, SEL is more prevalent in the MMER literature \citep{paper01, paper07}. Advanced techniques for decision-level fusion include late fusion subtask merging (LFSM), which integrates audio and lyrics for multimodal emotion categorization \citep{huertas}, and stacking, which assigns distinct weights to each category or modality for better integration \citep{paper02, paper08}. Recent advancements have focused on improving decision-level fusion by employing neural networks for weighting outputs. However, this approach has shown limited efficacy \cite{Hussain-2024}. Instead, stacking methods that combine the outputs of multiple models without fully fusing them have been shown to enhance performance \citep{paper10}.

To obtain unimodal emotion predictions, whether as final output or as input to multimodal models, various methods have been used. For audio, earlier studies have used SVMs, random forests, and logistic regression \citep{2008_1,2008_2,2015_1,Huang}, while nowadays most audio-based MER is done using deep learning methods such as CNNs and RNNs. Others have experimented with combining a CNN and an RNN in a single model \cite{2017-1}. Continuous audio-based MER is often done by a combination of CNNs and LSTMs \citep{paper10,paper12,paper06}. Furthermore, bidirectional LSTMs (BiLSTMs) with attention layers have been used \citep{paper10}. For audio classification, CNNs have been found to outperform BiGRU models \cite{Sujeesha}. Some researchers have leveraged transfer learning by using pretrained models on large-scale audio datasets, which are then fine-tuned for specific emotion recognition tasks \citep{paper07,paper19}. Building on this, fusion models have combined CNNs for SF extraction and LSTMs for TF extraction to enhance feature representation \citep{paper10,paper07}, while convolutional autoencoders with frequency and time masking techniques have been used to emphasize important features and reduce noise \citep{paper15}. Researchers have also adopted self-attention mechanisms in RNNs which dynamically weigh the importance of different audio segments \citep{paper01}, and end-to-end learning approaches like SampleCNN to process raw audio signals directly using very small filters \citep{paper15}. Additionally, they have incorporated multiview neural networks to integrate multiple perspectives of the audio signal \citep{paper10}.

Although audio has thus far been the most prominent modality in MER, other modalities are gaining popularity. Earlier studies on lyric-based MER largely utilized traditional classifiers such as SVMs \citep{2008_1,2008_2,2015_1}. However, these methods often fall short in capturing the deeper contextual and semantic nuances of lyrics, which are crucial for accurate emotion recognition \citep{paper12}. More recent research in lyric-based MER has shifted to deep learning models (Table \ref{tab:full_summary}). CNNs, RNNs, and LSTMs are utilized for their ability to handle sequential data and maintain context over long text spans. These models identify and retain emotional indicators spread across multiple lines of lyrics. Neural networks such as the restricted Boltzmann machine (RBM) are used for unsupervised feature learning for lyric classification \citep{2019_01}. Additionally, the BERT model, with its transformer-based architecture, is used to understand the bidirectional context of words, enhancing the accuracy of emotion detection in lyrics \citep{paper12,paper13}. To our knowledge, there exist no works using Approach 2-B with video. MIDI has also been used for training MER models using symbolic features derived from it. Notably, some studies have employed an SVM as the MER model \cite{2017-2}, while others have utilized an RNN, specifically a BiGRU \cite{paper11}. Physiological signals such as EEG have been used with an SVM model \cite{2017-2} and electrodermal activity with a CNN for MER \cite{EDA_music}. We are not aware of studies that have used textual data and metadata/contextual data for unimodal emotion recognition. 

\subsubsection{Emotion Prediction by Model-Level Fusion}
Model-level fusion, also referred to as middle-level or intermediate fusion, involves utilizing embeddings (Approach 2-A) or predictions derived from modality-specific models (Approach 2-B) to train a final emotion prediction model. This bridges feature-level and decision-level fusion, offering a balance between combining raw features and final predictions. Techniques for model-level fusion include combining embeddings generated by modality-specific models, such as those for audio and lyrics, into a shared representation for emotion recognition. Additionally, methods like the Hough voting mechanism within the Hough forest model have been explored, effectively integrating cues from multiple modalities to predict music emotions \citep{2015_1}. This allows modalities to contribute jointly while maintaining their distinct representations, thereby enhancing emotion prediction performance \citep{Delbouys}.

\subsubsection{Emotion Prediction by Cross-Modal Fusion}
Cross-modal feature fusion has emerged as a promising approach in MMER, gaining traction in more recent studies \cite{paper13, paper03, paper04}. This approach mainly aims at modeling and exploiting interaction between modalities at the feature level, effectively allowing the model to incorporate complementary information from different modalities. Unlike traditional fusion strategies, cross-modal fusion insists on direct processing of shared and unique aspects of each modality, fostering richer and more holistic representation.

Cross-modal processing typically involves aligning and integrating features from different modalities in a shared latent space, where relationships between modalities can be explored. Techniques such as attention mechanisms \cite{paper13} and graph-based methods are often used to model dependencies and interactions between modalities. For example, attention mechanisms dynamically weigh the importance of specific features in one modality based on their relevance to another, enhancing the model’s ability to prioritize meaningful interactions. Similarly, graph-based approaches construct cross-modal graphs to represent and learn complex relationships between modalities, facilitating a structured fusion process. Although graph-based approaches are not commonly applied in MMER, they have been widely used in other areas of multimodal research such as video understanding \cite{cross_graph_video}, and audio-visual event localization \cite{cross_graph_a_video_loc}.

\begin{footnotesize}
\setlength{\LTcapwidth}{\textwidth}
\begin{longtable}[!t]{>{\raggedright\arraybackslash}p{0.5cm}>{\raggedright\arraybackslash}p{0.5cm}>{\raggedright\arraybackslash}p{1.8cm}>{\raggedright\arraybackslash}p{2cm}>{\raggedright\arraybackslash}p{2cm}>{\centering\arraybackslash}p{1.2cm}>
{\raggedright\arraybackslash}p{2.42cm}}

\toprule
\multicolumn{1}{l}{\textbf{Reference}}&\multicolumn{1}{l}{\textbf{Year}} & \multicolumn{1}{l}{\textbf{Modalities}} & \multicolumn{1}{l}{\textbf{Datasets}} & \multicolumn{1}{l}{\textbf{Methods}} &\multicolumn{1}{c}{{\textbf{Approach}}} & {\textbf{Performance}} \\ \midrule
\endhead
\cite{2008_1} & 2008 &   Audio, Lyrics                     &  Own (last.fm + \newline LyricWiki)                                                      & SVM                                                                                    & \multicolumn{1}{c}{1}                                                                              &  Accuracy: 92.40\%                   \\ 
\midrule
\cite{2008_2} &  2008 &  Audio, Lyrics                        & Own                                                         & SVM                                                                                     & 2-A                                                                               & Accuracy:\newline 73.32\% Valence \newline 78.03\% Arousal                                  \\ \midrule
\cite{mirex-like} & 2013&   Audio, Lyrics, MIDI                         &  Own                                                         & SVM                                                                                     &\multicolumn{1}{c}{1}                                                                            & Accuracy: 61.20\%                                 \\ \midrule
\cite{2015_1} & 2015&   Audio, Lyrics                           &  Own                                                         & SVM                                                                                     & 1                                                                                & Accuracy: 59.9\%                                \\ 
\midrule
\cite{Huang} & 2016&   Audio, Lyrics                         &  Million Song DB                                                      &   SVM                                                                                      &2-A                                                                             & Accuracy: 72.18\%                       \\
\midrule

\cite{2017-1} & 2017&    Audio, Lyrics                         & Own                                                       & CNN+RNN, CNN                                                                                      & \multicolumn{1}{c}{2-A}                                                                                & Accuracy:  80.46\%                        \\
\midrule

\cite{2017-2} &  2017&  MIDI, EEG                         &  Own                                                       & SVM                                                                                      & 2-B                                                                              & Accuracy: 87.21\%                       \\
\midrule
\cite{Delbouys} & 2018 &   Audio, Lyrics                           & Own (Million Song DB $+$  Deezer Catalog)                                                      & CNN , LSTM                                                                                       & \multicolumn{1}{c}{2-A}                                                                                & R$^2$: 0.219 Valence\newline 0.232 Arousal                    \\ \midrule
\cite{2019_01} &  2019& Audio, Lyrics                          & Own                                                        & Bimodal Deep Auto Encoder                                                                                    & \multicolumn{1}{c}{1}                                                                                & Accuracy:\newline 79.9\% Pleasure \newline81.5\% Arousal \newline 67.7\% Dominance                    \\ \midrule
\cite{paper08}  & 2020                                                      & Audio, Lyrics                            & Own                                                         & LSTM , BERT                                                                                    & 2-B & Accuracy: 79.62\%                       \\ \midrule
\cite{paper07}   &2021                                                     & Audio, Lyrics                            & Million Songs DB& Multifeature Combined Classifier (CNN-LSTM Based)& 2-A                                                                                  & Accuracy: 78.2\%                         \\ \midrule
\cite{paper02} &2022& Audio, Video, Facial Expression & Own & CNN + OpenCV & 2-A & Accuracy: 74\% \\ \midrule

\cite{paper01}     &2022                                                 & Audio, Lyrics                            & MoodyLyrics                                                 & CNN , BERT                                                                                     & 2-B                                                                                           & Accuracy: 94.58\%                       \\ \midrule
\cite{paper10}                   &2022                                   & Audio, Lyrics                            & Own                                                         & CNN-LSTM                                                                                       & 2-A                                                                                      & Accuracy: 78.2\%                        \\ \midrule

\cite{paper12}    &2022                                                   & Audio, Videos                           & Own                                                         & CNN-LSTM                                                                                       & 2-A                         & Accuracy: 77.9\%                       \\ \midrule
\cite{paper13}       &2022                            &  Audio, Lyrics, Track Name, Artist       & Million Songs DB $+$ MoodyLyrics & CNN , BERT                                                                                     & 3                                                                                & R$^2$: 0.306 Valence\newline 0.311 Arousal \\ \midrule

\cite{paper09}    &2022                              & Audio, Lyrics                            & Own                                                         & CNN + NN                                                                                       & \multicolumn{1}{c}{2-B}                                                                                & Accuracy: 90.89\%                      \\ \midrule

\cite{EDA_music}& 2022&   Audio, Electrodermal Activity Signals                           & PMEmo + DEAP+ AMIGOS                                                       & RTCAN-1D         & \multicolumn{1}{c}{1}                                                                                & Accuracy:\newline79.68\% Valence\newline 83.76\% Arousal                       \\ \midrule

\cite{Sujeesha} & 2023&   Audio, Lyrics                        & MoodyLyrics                                                    & BiGRU + Attention                                                                                      & \multicolumn{1}{c}{2-A}                                                                                & Accuracy: 77.94\%                       \\ \midrule
 \cite{paper11}     &2023                                                   & Audio, MIDI                              & EMOPIA                                                      & BiGRU , CNN                                                                                    & 1                 & Accuracy: 69.2\%                     \\ \midrule
\cite{paper06}    &2023                                                    & Audio, Lyrics                            & Own (Indonesian) & CNN-LSTM , XLNet                                                                               & 1                                                                                      & Accuracy: 80.56\%                      \\ \midrule
\cite{paper03}     &2024                                                   & Audio, Lyrics                            & DEAM + FMA     & CNN , ALBERT                                                                                   & 3                   & Accuracy:\newline 49.68\% DEAM\newline 49.54\% FMA\\ \midrule
\cite{paper04}                   &2024                                     & Audio, Lyrics, Song Structure            & Own                                                         & OpenSMILE + BERT                                                                               & 3              & RMSE:\newline 0.162 Valence\newline 0.147 Arousal  \\ 
\bottomrule
\\
\caption{Summary of literature on MMER with performance.}
\label{tab:full_summary}
\end{longtable}
\end{footnotesize}

\section{Current Status and Future Directions}\label{sec:future}
The field of MMER has seen significant advancements in recent years. Nevertheless, there is still much room for further performance improvement (Table \ref{tab:full_summary}). To assist ongoing developments, we summarize the current trends (Section \ref{sec:trends}) and challenges (Section \ref{sec:challenges}), as well as potential future research directions (Section \ref{sec:directions}).

\subsection{Current Trends}\label{sec:trends}
Analyzing the recent literature, we observed some noticeable development trends in MMER, especially the growing availability of multimodal datasets and the shift from traditional machine learning methods to increasingly advanced multimodal deep learning methods.

\subsubsection{Multimodal Datasets}
With MMER gaining popularity, existing MER datasets have started to include data of different modalities, and new datasets are increasingly multimodal from the outset (Table \ref{tab:datasets}), including video, physiological signals, and textual data such as comments and metadata. Moreover, due to the dynamic nature of emotion in music, the field is now moving away from static processing, as it may not yield optimal and sufficiently detailed predictions for a given application. Therefore, datasets such as MERP \citep{MERP}, DEAM \citep{DEAM}, PMEmo \citep{PMEmo}, and  MuVi\citep{muVi} include dynamic annotation rather than static annotation. Also, the advancement of techniques such as RNN models has made dynamic processing more efficient, further accelerating the emergence of dynamically annotated datasets. One of the key driving factors in recent efforts to expand and create multimodal datasets is that they allow for the development of better emotion prediction methods. For example, in a study by Liu and Tan \cite{paper08}, the highest accuracies achieved by unimodal methods were 70.6\% for audio and 62.9\% for lyrics, while multimodal methods reached 79.2\%. Another study, by Chen and Li \cite{paper07}, reported 68.1\% and 74.2\% accuracy for audio and lyric classification, respectively, compared to 78.2\% accuracy for the multimodal model.

\subsubsection{Multimodal Methods} 
Over the years, the majority of MMER papers have considered audio and lyrics as the primary modalities (Table \ref{tab:full_summary}), and the developed methods have progressed from traditional machine learning techniques such as SVMs to more and more sophisticated deep learning models based on CNNs, RNNs (in particular LSTMs), and recently also Transformers \citep{paper07,paper10,paper12,paper06,Sujeesha}. In addition, various fusion strategies have been developed to effectively combine the two modalities. The highest accuracy of 94.58\% for classification to date has been achieved by employing a CNN for audio analysis, BERT for lyrics analysis, and using late fusion of the two \cite{paper01}. Other modalities such as  metadata, song structure, MIDI, and video have also been incorporated into MMER systems \citep{mirex-like,paper04,paper13}. A common approach is to combine audio with video, although the highest accuracy reported thus far is only 77.9\% \citep{paper12}. Modalities such as MIDI and physiological signals have as yet seen limited usage, with SVMs and CNNs being the most popular methods for prediction. The use of metadata has also received little attention to date. Processing of such data is typically done using models such as BERT and ALBERT \cite{paper13}.
    
Beyond traditional machine learning and deep learning models, some researchers have explored alternative technologies like the OpenSMILE\footnote{\url{https://www.audeering.com/research/opensmile/} (Accessed 15 November 2024).} toolkit, which can process a variety of modalities including audio, visual data, and physiological signals, offering a versatile approach to MMER. Another very important development highlighted in this survey is the use of cross-modal processing, which allows for more comprehensive and accurate emotion detection by enabling methods to process and correlate information from different sensory inputs simultaneously. Leveraging diverse modalities, cross-modal processing enhances the ability of MMER methods to understand complex emotional cues, leading to improved performance in various applications, from virtual assistants to mental health monitoring tools \cite{ISMIR-2010}.
    
\subsection{Current Challenges}\label{sec:challenges}
Despite recent advancements and the growing popularity of MMER, it is still quite far from achieving human-like performance (Table \ref{tab:full_summary}). One of the main challenges is the subjectivity of music experience. Different people can feel different emotions in a single song. Even the same person can feel different emotions for the same song depending on their mood \cite{mood_and_em}. According to research, emotion can be influenced by a variety of factors, including culture, genre preference, age, level of music expertise, gender, and even the weather. Therefore, some studies have considered the profile data of annotators and listeners when making predictions \citep{MERP}, and more recent works \cite{siTunes} even considered the time and weather during listening to the song when making predictions. Accurately classifying music using categorical emotion models is very difficult. In dimensional emotion models, particularly the widely used valence-arousal model, each quadrant has an extensive range of emotions (Fig.~\ref{fig:russells}), making precise annotation ambiguous and challenging. It might be beneficial to introduce a new emotion model specifically for the purpose of MER.

Another primary issue is the limited availability of datasets. The majority of datasets, such as DEAP, emoMV, 4Q \citep{deap,emoMV,4Q} do not include complete song audios due to copyright restrictions. Furthermore, the number of songs or song excerpts is very limited, ranging from 100s to 1,000s. Some more recent datasets including Audioset \citep{audioset}, MTG-Jamendo \citep{mtg}, MuSe \citep{MuSe}, and Music4all \citep{music4all} have ~17k, ~55k, 90k, and 109k songs, respectively (Table \ref{tab:datasets}), but they contain only a single modality (mostly audio only), making them unsuitable for MMER. Another limitation of current datasets is that they tend to focus on a single genre of music, resulting in a lack of diversity in the songs included. For instance, DEAM mainly includes rock and techno music, and MSD (Million Song Dataset) \citep{msd} consists primarily of pop music. Thus, there is an urgent need for larger and more diverse multimodal datasets for MMER. This would also open the door for the creation of state-of-the-art benchmarks for MMER, which are currently lacking.

A continuing problem in MER in general is the insufficient use of music theory and concepts, such as considering the mode the music is written in and loudness variations, in the process of recognizing emotions. It is known that concepts like tempo, pitch, and rhythm can convey emotion (Section \ref{TEM}), but thus far only a small number of studies have made use of these.

\subsection{Future Directions}\label{sec:directions}
To deal with dataset limitations, one option is to employ unsupervised learning methods such as autoencoders \citep{2019_01} as an alternative to the supervised approaches predominantly employed in prior studies. Unsupervised learning enables the utilization of custom unlabeled datasets, which researchers can generate using various APIs and tools. For example, the SoundCloud API\footnote{\url{https://developers.soundcloud.com/docs/api/guide} (Accessed 15 November 2024).} can facilitate audio data collection, Lyrics.ovh\footnote{\url{https://lyricsovh.docs.apiary.io} (Accessed 15 November 2024).} and MusixMatch\footnote{\url{https://developer.musixmatch.com} (Accessed 15 November 2024).} can provide lyrics data, and video data can be sourced using Python libraries such as PyTube\footnote{\url{https://pytube.io/} (Accessed 15 November 2024).}. This approach allows researchers to overcome reliance on preexisting datasets, offering greater flexibility and customization in data collection to meet specific research objectives.
Additionally, incorporating annotators from different cultures when creating datasets can provide a diverse perspective on music emotions. Another approach to overcome the dataset limitation problem is to use transfer learning. Although transfer learning is widely utilized and has produced excellent results in other domains such as computer vision \citep{transfer_learning}, it has seen very little use in the field of MER, and should be explored in the future. %As a result, this area remains unexplored and ought to be explored in the future to improve the study of MER.

Another potential future direction for MER is to incorporate different modalities other than only audio and lyrics \cite{2011-MER}. Researchers can acquire a better understanding of the listener's emotional response to music by combining physiological data like EEG signals, heart rate variability, and skin conductance. These physiological indicators offer a more objective assessment of emotional states, complementing personal judgements based on audio and lyrical content. MIDI is also an invaluable modality to incorporate. To comprehend the structural and expressive elements of a piece, one can analyze the precise symbolic information that MIDI provides about the music, such as note pitches, intervals, and dynamics. Furthermore, as noted in the previous section, involving music theory and concepts is a promising future direction. Integrating these many senses can result in a more complete and comprehensive understanding of emotions in music \cite{ISMIR-2010}. 

Real-time MER is another promising future direction with many possible applications. Currently, many MER models are neither user-friendly nor optimized for quick execution, demanding GPUs and considerable processing time. However, the advancement of real-time technology has the potential to transform multiple sectors. Real-time emotion detection in therapeutic settings can improve emotional regulation techniques and mental health treatments by giving instant feedback to therapists and patients. Real-time analysis could help mood guide playlist systems improve user experience by dynamically changing song choices to better fit the listener's current emotional state. Furthermore, real-time emotion identification could be used in commercial applications to deliver more effective and personalized marketing content, such as targeted advertising \citep{ext_3}. Real-time emotion recognition in music has great potential for both private and public applications as technology develops and these models become more affordable and effective.

Creating real-time MMER applications is quite challenging, primarily due to high computational demands. Combining and synchronizing data streams like audio, video, and text requires advanced algorithms and significant processing power \cite{ext_1}. This complexity often hampers real-time processing. Additionally, the substantial resource demands such as increased memory, processing power, and energy consumption pose further difficulties. This is particularly problematic for portable or embedded systems, where resources are limited. As a result, implementing real-time MMER on less powerful devices can be impractical, hindering broader applications \cite{ext_2}. However, these can be mitigated by adopting efficient data fusion methods. Developing lightweight algorithms that effectively combine multiple modalities with minimal computational overhead is a key strategy. Additionally, leveraging parallel processing techniques or utilizing specialized hardware, such as GPUs, can significantly enhance the system's ability to handle multiple data streams simultaneously without compromising speed. These approaches help maintain the real-time functionality of the system while addressing the inherent computational and resource demands.

\section{Conclusion}\label{conclusion}
MMER is an upcoming and developing field in the area of music information retrieval. Although unimodal MER has achieved exceptional performances, MMER has not reached human-like accuracy yet. From the studies surveyed in this paper, we conclude that the majority of MMER methods are limited to using audio and lyrics as the primary modalities. Incorporating other modalities such as video, MIDI, physiological signals, metadata and other (con)textural data may provide additional information and improve emotion recognition performance. Moreover, as most works employ machine learning and deep learning methods to train emotion prediction models, primary factors limiting progress in MMER are the absence of a precise emotional model, limited datasets, and a lack of state-of-the-art benchmarks. Thus, future research should focus on the integration of a wider variety of modalities, the exploration of new deep learning architectures, and the development of comprehensive evaluation frameworks. Notwithstanding current challenges and limitations, MMER holds significant promise. As the field continues to grow, MMER will not only bridge the gap between human-like emotional understanding and machine perception, but also introduce new opportunities for enhancing user experiences and emotional intelligence in technology, revolutionizing human-computer interaction, multimedia content production, medical treatment, recommendation systems, and commercial applications.

% \section{Authors and Affiliations}

% Each author must be defined separately for accurate metadata
% identification.  As an exception, multiple authors may share one
% affiliation. Authors' names should not be abbreviated; use full first
% names wherever possible. Include authors' e-mail addresses whenever
% possible.

% Grouping authors' names or e-mail addresses, or providing an ``e-mail
% alias,'' as shown below, is not acceptable:
% \begin{verbatim}
%   \author{Brooke Aster, David Mehldau}
%   \email{dave,judy,steve@university.edu}
%   \email{firstname.lastname@phillips.org}
% \end{verbatim}

% The \verb|authornote| and \verb|authornotemark| commands allow a note
% to apply to multiple authors --- for example, if the first two authors
% of an article contributed equally to the work.

% If your author list is lengthy, you must define a shortened version of
% the list of authors to be used in the page headers, to prevent
% overlapping text. The following command should be placed just after
% the last \verb|\author{}| definition:
% \begin{verbatim}
%   \renewcommand{\shortauthors}{McCartney, et al.}
% \end{verbatim}
% Omitting this command will force the use of a concatenated list of all
% of the authors' names, which may result in overlapping text in the
% page headers.

% The article template's documentation, available at
% \url{https://www.acm.org/publications/proceedings-template}, has a
% complete explanation of these commands and tips for their effective
% use.

% Note that authors' addresses are mandatory for journal articles.

\bibliographystyle{ACM-Reference-Format}
\bibliography{sample-base}


\begin{thebibliography}{119}

%%% the Web address.

\ifx \showCODEN    \undefined \def \showCODEN     #1{\unskip}     \fi
\ifx \showDOI      \undefined \def \showDOI       #1{#1}\fi
\ifx \showISBNx    \undefined \def \showISBNx     #1{\unskip}     \fi
\ifx \showISBNxiii \undefined \def \showISBNxiii  #1{\unskip}     \fi
\ifx \showISSN     \undefined \def \showISSN      #1{\unskip}     \fi
\ifx \showLCCN     \undefined \def \showLCCN      #1{\unskip}     \fi
\ifx \shownote     \undefined \def \shownote      #1{#1}          \fi
\ifx \showarticletitle \undefined \def \showarticletitle #1{#1}   \fi
\ifx \showURL      \undefined \def \showURL       {\relax}        \fi
% invisible to TeX
\providecommand\bibfield[2]{#2}
\providecommand\bibinfo[2]{#2}
\providecommand\natexlab[1]{#1}
\providecommand\showeprint[2][]{arXiv:#2}

\bibitem[Akiki and Burghardt(2021)]%
        {MuSe}
\bibfield{author}{\bibinfo{person}{Christopher Akiki} {and} \bibinfo{person}{Manuel Burghardt}.} \bibinfo{year}{2021}\natexlab{}.
\newblock \showarticletitle{{MuSe}: The musical sentiment dataset}.
\newblock \bibinfo{journal}{\emph{Journal of Open Humanities Data}}  \bibinfo{volume}{7} (\bibinfo{year}{2021}), \bibinfo{pages}{10}.
\newblock
\urldef\tempurl%
\url{https://doi.org/10.5334/johd.33}
\showDOI{\tempurl}


\bibitem[Aljanaki et~al\mbox{.}(2016)]%
        {emotify}
\bibfield{author}{\bibinfo{person}{Anna Aljanaki}, \bibinfo{person}{Frans Wiering}, {and} \bibinfo{person}{Remco~C. Veltkamp}.} \bibinfo{year}{2016}\natexlab{}.
\newblock \showarticletitle{Studying emotion induced by music through a crowdsourcing game}.
\newblock \bibinfo{journal}{\emph{Information Processing \& Management}} \bibinfo{volume}{52}, \bibinfo{number}{1} (\bibinfo{year}{2016}), \bibinfo{pages}{115–128}.
\newblock
\showISSN{0306-4573}
\urldef\tempurl%
\url{https://doi.org/10.1016/j.ipm.2015.03.004}
\showDOI{\tempurl}


\bibitem[Aljanaki et~al\mbox{.}(2017)]%
        {DEAM}
\bibfield{author}{\bibinfo{person}{Anna Aljanaki}, \bibinfo{person}{Yi-Hsuan Yang}, {and} \bibinfo{person}{Mohammad Soleymani}.} \bibinfo{year}{2017}\natexlab{}.
\newblock \showarticletitle{Developing a benchmark for emotional analysis of music}.
\newblock \bibinfo{journal}{\emph{PLOS ONE}} \bibinfo{volume}{12}, \bibinfo{number}{3} (\bibinfo{year}{2017}), \bibinfo{pages}{e0173392}.
\newblock
\showISSN{1932-6203}
\urldef\tempurl%
\url{https://doi.org/10.1371/journal.pone.0173392}
\showDOI{\tempurl}


\bibitem[Baltrušaitis et~al\mbox{.}(2017)]%
        {ext_1}
\bibfield{author}{\bibinfo{person}{Tadas Baltrušaitis}, \bibinfo{person}{Chaitanya Ahuja}, {and} \bibinfo{person}{Louis-Philippe Morency}.} \bibinfo{year}{2017}\natexlab{}.
\newblock \bibinfo{title}{Multimodal Machine Learning: A Survey and Taxonomy}.
\newblock
\newblock
\showeprint[arxiv]{1705.09406}~[cs.LG]
\urldef\tempurl%
\url{https://arxiv.org/abs/1705.09406}
\showURL{%
\tempurl}


\bibitem[Bao et~al\mbox{.}(2011)]%
        {pyegg}
\bibfield{author}{\bibinfo{person}{Forrest~Sheng Bao}, \bibinfo{person}{Xin Liu}, {and} \bibinfo{person}{Christina Zhang}.} \bibinfo{year}{2011}\natexlab{}.
\newblock \showarticletitle{{PyEEG}: An open source {Python} module for {EEG/MEG} feature extraction}.
\newblock \bibinfo{journal}{\emph{Computational Intelligence and Neuroscience}} \bibinfo{volume}{2011}, \bibinfo{number}{1} (\bibinfo{year}{2011}), \bibinfo{pages}{406391}.
\newblock
\urldef\tempurl%
\url{https://doi.org/10.1155/2011/406391}
\showDOI{\tempurl}


\bibitem[Barrett(2006)]%
        {barretts}
\bibfield{author}{\bibinfo{person}{Lisa~Feldman Barrett}.} \bibinfo{year}{2006}\natexlab{}.
\newblock \showarticletitle{Solving the emotion paradox: Categorization and the experience of emotion}.
\newblock \bibinfo{journal}{\emph{Personality and Social Psychology Review}} \bibinfo{volume}{10}, \bibinfo{number}{1} (\bibinfo{year}{2006}), \bibinfo{pages}{20--46}.
\newblock
\urldef\tempurl%
\url{https://doi.org/10.1207/s15327957pspr1001\_2}
\showDOI{\tempurl}
\newblock
\shownote{PMID: 16430327}.


\bibitem[Bertin-Mahieux et~al\mbox{.}(2011)]%
        {msd}
\bibfield{author}{\bibinfo{person}{Thierry Bertin-Mahieux}, \bibinfo{person}{Daniel~P.W. Ellis}, \bibinfo{person}{Brian Whitman}, {and} \bibinfo{person}{Paul Lamere}.} \bibinfo{year}{2011}\natexlab{}.
\newblock \showarticletitle{The million song dataset}. In \bibinfo{booktitle}{\emph{International Society for Music Information Retrieval Conference ({ISMIR})}}. \bibinfo{publisher}{International Society for Music Information Retrieval}, \bibinfo{address}{Miami, Florida, USA}, \bibinfo{pages}{591--596}.
\newblock
\urldef\tempurl%
\url{https://doi.org/10.7916/D8NZ8J07}
\showDOI{\tempurl}


\bibitem[Bogdanov et~al\mbox{.}(2022)]%
        {musAV}
\bibfield{author}{\bibinfo{person}{Dmitry Bogdanov}, \bibinfo{person}{Xabier Lizarraga-Seijas}, \bibinfo{person}{Pablo Alonso-Jiménez}, {and} \bibinfo{person}{Xavier Serra}.} \bibinfo{year}{2022}\natexlab{}.
\newblock \showarticletitle{{MusAV}: A dataset of relative arousal-valence annotations for validation of audio models}. In \bibinfo{booktitle}{\emph{International Society for Music Information Retrieval Conference (ISMIR)}}. \bibinfo{publisher}{International Society for Music Information Retrieval}, \bibinfo{address}{Bengaluru, India}, \bibinfo{pages}{650--658}.
\newblock
\urldef\tempurl%
\url{https://doi.org/10.5281/zenodo.7316746}
\showDOI{\tempurl}


\bibitem[Bogdanov et~al\mbox{.}(2019)]%
        {mtg}
\bibfield{author}{\bibinfo{person}{Dmitry Bogdanov}, \bibinfo{person}{Minz Won}, \bibinfo{person}{Philip Tovstogan}, \bibinfo{person}{Alastair Porter}, {and} \bibinfo{person}{Xavier Serra}.} \bibinfo{year}{2019}\natexlab{}.
\newblock \showarticletitle{The {MTG-Jamendo} dataset for automatic music tagging}. In \bibinfo{booktitle}{\emph{International Conference on Machine Learning}}. \bibinfo{publisher}{Semantic Scholar}, \bibinfo{address}{Long Beach, California, USA}, \bibinfo{pages}{1--3}.
\newblock
\urldef\tempurl%
\url{https://api.semanticscholar.org/CorpusID:196187495}
\showURL{%
\tempurl}


\bibitem[Burke(2017)]%
        {audio_F}
\bibfield{author}{\bibinfo{person}{Cibele~Maia Burke}.} \bibinfo{year}{2017}\natexlab{}.
\newblock \showarticletitle{A comparative study of perspectives in musical structural features and emotional stimuli}. In \bibinfo{booktitle}{\emph{Honors Theses}}. \bibinfo{publisher}{Eastern Kentucky University}, \bibinfo{address}{Kentucky, USA}, \bibinfo{pages}{1--21}.
\newblock
\urldef\tempurl%
\url{https://api.semanticscholar.org/CorpusID:67822675}
\showURL{%
\tempurl}


\bibitem[Carr et~al\mbox{.}(2023)]%
        {articulation}
\bibfield{author}{\bibinfo{person}{Nathan~R. Carr}, \bibinfo{person}{Kirk~N. Olsen}, {and} \bibinfo{person}{William~Forde Thompson}.} \bibinfo{year}{2023}\natexlab{}.
\newblock \showarticletitle{The perceptual and emotional consequences of articulation in music}.
\newblock \bibinfo{journal}{\emph{Music Perception}} \bibinfo{volume}{40}, \bibinfo{number}{3} (\bibinfo{year}{2023}), \bibinfo{pages}{202--219}.
\newblock
\showISSN{0730-7829}
\urldef\tempurl%
\url{https://doi.org/10.1525/mp.2023.40.3.202}
\showDOI{\tempurl}


\bibitem[Chaturvedi et~al\mbox{.}(2021)]%
        {psycho}
\bibfield{author}{\bibinfo{person}{Vybhav Chaturvedi}, \bibinfo{person}{Arman~Beer Kaur}, \bibinfo{person}{Vedansh Varshney}, \bibinfo{person}{Anupam Garg}, \bibinfo{person}{Gurpal~Singh Chhabra}, {and} \bibinfo{person}{Munish Kumar}.} \bibinfo{year}{2021}\natexlab{}.
\newblock \showarticletitle{Music mood and human emotion recognition based on physiological signals: A systematic review}.
\newblock \bibinfo{journal}{\emph{Multimedia Systems}}  \bibinfo{volume}{28} (\bibinfo{year}{2021}), \bibinfo{pages}{21--44}.
\newblock
\urldef\tempurl%
\url{https://api.semanticscholar.org/CorpusID:234871442}
\showURL{%
\tempurl}


\bibitem[Chen and Li(2020)]%
        {paper07}
\bibfield{author}{\bibinfo{person}{Changfeng Chen} {and} \bibinfo{person}{Qiang Li}.} \bibinfo{year}{2020}\natexlab{}.
\newblock \showarticletitle{A multimodal music emotion classification method based on multifeature combined network classifier}.
\newblock \bibinfo{journal}{\emph{Mathematical Problems in Engineering}} \bibinfo{volume}{2020}, \bibinfo{number}{1} (\bibinfo{year}{2020}), \bibinfo{pages}{4606027}.
\newblock
\urldef\tempurl%
\url{https://doi.org/10.1155/2020/4606027}
\showDOI{\tempurl}


\bibitem[Chen(2022)]%
        {paper12}
\bibfield{author}{\bibinfo{person}{Wenwen Chen}.} \bibinfo{year}{2022}\natexlab{}.
\newblock \showarticletitle{A novel long short-term memory network model for multimodal music emotion analysis in affective computing}.
\newblock \bibinfo{journal}{\emph{Journal of Applied Science and Engineering}} \bibinfo{volume}{26}, \bibinfo{number}{3} (\bibinfo{year}{2022}), \bibinfo{pages}{367--376}.
\newblock
\showISSN{1560-6686}
\urldef\tempurl%
\url{https://doi.org/10.6180/jase.202303_26(3).0008}
\showDOI{\tempurl}


\bibitem[Chen et~al\mbox{.}(2015)]%
        {AMG1608}
\bibfield{author}{\bibinfo{person}{Yu-An Chen}, \bibinfo{person}{Yi-Hsuan Yang}, \bibinfo{person}{Ju-Chiang Wang}, {and} \bibinfo{person}{Homer Chen}.} \bibinfo{year}{2015}\natexlab{}.
\newblock \showarticletitle{The {AMG1608} dataset for music emotion recognition}. In \bibinfo{booktitle}{\emph{IEEE International Conference on Acoustics, Speech and Signal Processing (ICASSP)}}. \bibinfo{publisher}{IEEE}, \bibinfo{address}{South Brisbane, Queensland, Australia}, \bibinfo{pages}{693--697}.
\newblock
\urldef\tempurl%
\url{https://doi.org/10.1109/ICASSP.2015.7178058}
\showDOI{\tempurl}


\bibitem[Choi et~al\mbox{.}(2017)]%
        {choi}
\bibfield{author}{\bibinfo{person}{Keunwoo Choi}, \bibinfo{person}{Gy{\"{o}}rgy Fazekas}, \bibinfo{person}{Mark~B. Sandler}, {and} \bibinfo{person}{Kyunghyun Cho}.} \bibinfo{year}{2017}\natexlab{}.
\newblock \showarticletitle{Convolutional recurrent neural networks for music classification}. In \bibinfo{booktitle}{\emph{IEEE International Conference on Acoustics, Speech and Signal Processing (ICASSP)}}. \bibinfo{publisher}{IEEE}, \bibinfo{address}{New Orleans, Louisiana, USA}, \bibinfo{pages}{2392--2396}.
\newblock
\urldef\tempurl%
\url{https://doi.org/10.1109/ICASSP.2017.7952585}
\showDOI{\tempurl}


\bibitem[Chua et~al\mbox{.}(2022)]%
        {muVi}
\bibfield{author}{\bibinfo{person}{Phoebe Chua}, \bibinfo{person}{Dimos Makris}, \bibinfo{person}{Dorien Herremans}, \bibinfo{person}{Gemma Roig}, {and} \bibinfo{person}{Kat Agres}.} \bibinfo{year}{2022}\natexlab{}.
\newblock \bibinfo{title}{Predicting emotion from music videos: Exploring the relative contribution of visual and auditory information to affective responses}.
\newblock
\newblock
\showeprint[arxiv]{2202.10453}
\urldef\tempurl%
\url{https://doi.org/10.48550/arXiv.2202.10453}
\showURL{%
\tempurl}


\bibitem[Cui et~al\mbox{.}(2022)]%
        {EEG_survey}
\bibfield{author}{\bibinfo{person}{Xu Cui}, \bibinfo{person}{Yongrong Wu}, \bibinfo{person}{Jipeng Wu}, \bibinfo{person}{Zhiyu You}, \bibinfo{person}{Jianbing Xiahou}, {and} \bibinfo{person}{Menglin Ouyang}.} \bibinfo{year}{2022}\natexlab{}.
\newblock \showarticletitle{A review: Music-emotion recognition and analysis based on {EEG} signals}.
\newblock \bibinfo{journal}{\emph{Frontiers in Neuroinformatics}}  \bibinfo{volume}{16} (\bibinfo{year}{2022}), \bibinfo{pages}{1--17}.
\newblock
\showISSN{1662-5196}
\urldef\tempurl%
\url{https://doi.org/10.3389/fninf.2022.997282}
\showDOI{\tempurl}


\bibitem[Dakshina and Sridhar(2014)]%
        {unigram}
\bibfield{author}{\bibinfo{person}{K. Dakshina} {and} \bibinfo{person}{Rajeswari Sridhar}.} \bibinfo{year}{2014}\natexlab{}.
\newblock \showarticletitle{{LDA} based emotion recognition from lyrics}. In \bibinfo{booktitle}{\emph{Advanced Computing, Networking and Informatics}}, Vol.~\bibinfo{volume}{1}. \bibinfo{publisher}{Springer International Publishing}, \bibinfo{address}{Cham, Switzerland}, \bibinfo{pages}{187--194}.
\newblock
\showISBNx{978-3-319-07353-8}
\urldef\tempurl%
\url{https://doi.org/10.1007/978-3-319-07353-8_22}
\showDOI{\tempurl}


\bibitem[Delbouys et~al\mbox{.}(2018)]%
        {Delbouys}
\bibfield{author}{\bibinfo{person}{Rémi Delbouys}, \bibinfo{person}{Romain Hennequin}, \bibinfo{person}{Francesco Piccoli}, \bibinfo{person}{Jimena Royo-Letelier}, {and} \bibinfo{person}{Manuel Moussallam}.} \bibinfo{year}{2018}\natexlab{}.
\newblock \bibinfo{title}{Music mood detection based on audio and lyrics with deep neural net}.
\newblock
\newblock
\showeprint[arxiv]{1809.07276}
\urldef\tempurl%
\url{https://doi.org/10.48550/arXiv.1809.07276}
\showURL{%
\tempurl}


\bibitem[Ding et~al\mbox{.}(2016)]%
        {cnn_v_2}
\bibfield{author}{\bibinfo{person}{Wan Ding}, \bibinfo{person}{Mingyu Xu}, \bibinfo{person}{Dongyan Huang}, \bibinfo{person}{Weisi Lin}, \bibinfo{person}{Minghui Dong}, \bibinfo{person}{Xinguo Yu}, {and} \bibinfo{person}{Haizhou Li}.} \bibinfo{year}{2016}\natexlab{}.
\newblock \showarticletitle{Audio and face video emotion recognition in the wild using deep neural networks and small datasets}. In \bibinfo{booktitle}{\emph{ACM International Conference on Multimodal Interaction (ICMI)}}. \bibinfo{publisher}{Association for Computing Machinery}, \bibinfo{address}{New York, NY, USA}, \bibinfo{pages}{506--513}.
\newblock
\showISBNx{9781450345569}
\urldef\tempurl%
\url{https://doi.org/10.1145/2993148.2997637}
\showDOI{\tempurl}


\bibitem[Dogaru et~al\mbox{.}(2024)]%
        {music_in_ads}
\bibfield{author}{\bibinfo{person}{Isabela Dogaru}, \bibinfo{person}{Adrian Furnham}, {and} \bibinfo{person}{Alastair McClelland}.} \bibinfo{year}{2024}\natexlab{}.
\newblock \showarticletitle{Understanding how the presence of music in advertisements influences consumer behaviour}.
\newblock \bibinfo{journal}{\emph{Acta Psychologica}}  \bibinfo{volume}{248} (\bibinfo{year}{2024}), \bibinfo{pages}{104333}.
\newblock
\showISSN{0001-6918}
\urldef\tempurl%
\url{https://doi.org/10.1016/j.actpsy.2024.104333}
\showDOI{\tempurl}


\bibitem[Edmonds and Sedoc(2021)]%
        {edmonds}
\bibfield{author}{\bibinfo{person}{Darren Edmonds} {and} \bibinfo{person}{Jo{\~a}o Sedoc}.} \bibinfo{year}{2021}\natexlab{}.
\newblock \showarticletitle{Multi-emotion classification for song lyrics}. In \bibinfo{booktitle}{\emph{Workshop on Computational Approaches to Subjectivity, Sentiment and Social Media Analysis}}. \bibinfo{publisher}{Association for Computational Linguistics}, \bibinfo{address}{Kerrville, Texas, USA}, \bibinfo{pages}{221--235}.
\newblock
\urldef\tempurl%
\url{https://aclanthology.org/2021.wassa-1.24}
\showURL{%
\tempurl}


\bibitem[Eerola and Vuoskoski(2011)]%
        {soundtracks}
\bibfield{author}{\bibinfo{person}{Tuomas Eerola} {and} \bibinfo{person}{Jonna~K. Vuoskoski}.} \bibinfo{year}{2011}\natexlab{}.
\newblock \showarticletitle{A comparison of the discrete and dimensional models of emotion in music}.
\newblock \bibinfo{journal}{\emph{Psychology of Music}} \bibinfo{volume}{39}, \bibinfo{number}{1} (\bibinfo{year}{2011}), \bibinfo{pages}{18–49}.
\newblock
\showISSN{0305-7356, 1741-3087}
\urldef\tempurl%
\url{https://doi.org/10.1177/0305735610362821}
\showDOI{\tempurl}


\bibitem[Ekman(1971)]%
        {ekman1971universals}
\bibfield{author}{\bibinfo{person}{Paul Ekman}.} \bibinfo{year}{1971}\natexlab{}.
\newblock \showarticletitle{Universals and cultural differences in facial expressions of emotion}. In \bibinfo{booktitle}{\emph{Nebraska Symposium on Motivation}}. \bibinfo{publisher}{University of Nebraska Press}, \bibinfo{address}{Nebraska, USA}, \bibinfo{pages}{207–283}.
\newblock
\urldef\tempurl%
\url{https://psycnet.apa.org/record/1973-11154-001}
\showURL{%
\tempurl}


\bibitem[Fu et~al\mbox{.}(2011)]%
        {Zhouyu}
\bibfield{author}{\bibinfo{person}{Zhouyu Fu}, \bibinfo{person}{Guojun Lu}, \bibinfo{person}{Kai Ting}, {and} \bibinfo{person}{Dengsheng Zhang}.} \bibinfo{year}{2011}\natexlab{}.
\newblock \showarticletitle{A survey of audio-based music classification and annotation}.
\newblock \bibinfo{journal}{\emph{IEEE Transactions on Multimedia}} \bibinfo{volume}{13}, \bibinfo{number}{2} (\bibinfo{year}{2011}), \bibinfo{pages}{303--319}.
\newblock
\urldef\tempurl%
\url{https://doi.org/10.1109/TMM.2010.2098858}
\showDOI{\tempurl}


\bibitem[Gemmeke et~al\mbox{.}(2017)]%
        {audioset}
\bibfield{author}{\bibinfo{person}{Jort~F. Gemmeke}, \bibinfo{person}{Daniel P.~W. Ellis}, \bibinfo{person}{Dylan Freedman}, \bibinfo{person}{Aren Jansen}, \bibinfo{person}{Wade Lawrence}, \bibinfo{person}{R.~Channing Moore}, \bibinfo{person}{Manoj Plakal}, {and} \bibinfo{person}{Marvin Ritter}.} \bibinfo{year}{2017}\natexlab{}.
\newblock \showarticletitle{{Audio Set}: An ontology and human-labeled dataset for audio events}. In \bibinfo{booktitle}{\emph{IEEE International Conference on Acoustics, Speech and Signal Processing (ICASSP)}}. \bibinfo{publisher}{IEEE}, \bibinfo{address}{New Orleans, Louisiana, USA}, \bibinfo{pages}{776--780}.
\newblock
\urldef\tempurl%
\url{https://doi.org/10.1109/ICASSP.2017.7952261}
\showDOI{\tempurl}


\bibitem[G\'{o}mez-Ca\~{n}\'{o}n et~al\mbox{.}(2022)]%
        {Tompa}
\bibfield{author}{\bibinfo{person}{Juan~Sebasti\'{a}n G\'{o}mez-Ca\~{n}\'{o}n}, \bibinfo{person}{Nicol\'{a}s Guti\'{e}rrez-P\'{a}ez}, \bibinfo{person}{Lorenzo Porcaro}, \bibinfo{person}{Alastair Porter}, \bibinfo{person}{Estefan\'{\i}a Cano}, \bibinfo{person}{Perfecto Herrera-Boyer}, \bibinfo{person}{Aggelos Gkiokas}, \bibinfo{person}{Patricia Santos}, \bibinfo{person}{Davinia Hern\'{a}ndez-Leo}, \bibinfo{person}{Casper Karreman}, {and} \bibinfo{person}{Emilia G\'{o}mez}.} \bibinfo{year}{2022}\natexlab{}.
\newblock \showarticletitle{{TROMPA-MER}: An open dataset for personalized music emotion recognition}.
\newblock \bibinfo{journal}{\emph{Journal of Intelligent Information Systems}} \bibinfo{volume}{60}, \bibinfo{number}{2} (\bibinfo{year}{2022}), \bibinfo{pages}{549–570}.
\newblock
\showISSN{0925-9902}
\urldef\tempurl%
\url{https://doi.org/10.1007/s10844-022-00746-0}
\showDOI{\tempurl}


\bibitem[Grigorev et~al\mbox{.}(2024)]%
        {siTunes}
\bibfield{author}{\bibinfo{person}{Vadim Grigorev}, \bibinfo{person}{Jiayu Li}, \bibinfo{person}{Weizhi Ma}, \bibinfo{person}{Zhiyu He}, \bibinfo{person}{Min Zhang}, \bibinfo{person}{Yiqun Liu}, \bibinfo{person}{Ming Yan}, {and} \bibinfo{person}{Ji Zhang}.} \bibinfo{year}{2024}\natexlab{}.
\newblock \showarticletitle{{SiTunes}: A situational music recommendation dataset with physiological and psychological signals}. In \bibinfo{booktitle}{\emph{Conference on Human Information Interaction and Retrieval (CHIIR)}}. \bibinfo{publisher}{Association for Computing Machinery}, \bibinfo{address}{New York, NY, USA}, \bibinfo{pages}{417–421}.
\newblock
\showISBNx{9798400704345}
\urldef\tempurl%
\url{https://doi.org/10.1145/3627508.3638343}
\showDOI{\tempurl}


\bibitem[Han et~al\mbox{.}(2022)]%
        {2022-Overall}
\bibfield{author}{\bibinfo{person}{Donghong Han}, \bibinfo{person}{Yanru Kong}, \bibinfo{person}{Han Jiayi}, {and} \bibinfo{person}{Guoren Wang}.} \bibinfo{year}{2022}\natexlab{}.
\newblock \showarticletitle{A survey of music emotion recognition}.
\newblock \bibinfo{journal}{\emph{Frontiers of Computer Science}}  \bibinfo{volume}{16} (\bibinfo{year}{2022}), \bibinfo{pages}{166335}.
\newblock
\urldef\tempurl%
\url{https://doi.org/10.1007/s11704-021-0569-4}
\showDOI{\tempurl}


\bibitem[Han et~al\mbox{.}(2023)]%
        {paper19}
\bibfield{author}{\bibinfo{person}{Xiao Han}, \bibinfo{person}{Fuyang Chen}, {and} \bibinfo{person}{Junrong Ban}.} \bibinfo{year}{2023}\natexlab{}.
\newblock \showarticletitle{Music emotion recognition based on a neural network with an inception-gru residual structure}.
\newblock \bibinfo{journal}{\emph{Electronics}} \bibinfo{volume}{12}, \bibinfo{number}{4} (\bibinfo{year}{2023}), \bibinfo{pages}{978}.
\newblock
\showISSN{2079-9292}
\urldef\tempurl%
\url{https://doi.org/10.3390/electronics12040978}
\showDOI{\tempurl}


\bibitem[He and Ferguson(2022)]%
        {paper15}
\bibfield{author}{\bibinfo{person}{Na He} {and} \bibinfo{person}{Sam Ferguson}.} \bibinfo{year}{2022}\natexlab{}.
\newblock \showarticletitle{Music emotion recognition based on segment-level two-stage learning}.
\newblock \bibinfo{journal}{\emph{International Journal of Multimedia Information Retrieval}} \bibinfo{volume}{11}, \bibinfo{number}{3} (\bibinfo{year}{2022}), \bibinfo{pages}{383–394}.
\newblock
\urldef\tempurl%
\url{https://doi.org/10.1007/s13735-022-00230-z}
\showDOI{\tempurl}


\bibitem[Higuchi(1988)]%
        {FD}
\bibfield{author}{\bibinfo{person}{T. Higuchi}.} \bibinfo{year}{1988}\natexlab{}.
\newblock \showarticletitle{Approach to an irregular time series on the basis of the fractal theory}.
\newblock \bibinfo{journal}{\emph{Physica D: Nonlinear Phenomena}} \bibinfo{volume}{31}, \bibinfo{number}{2} (\bibinfo{year}{1988}), \bibinfo{pages}{277–283}.
\newblock
\showISSN{0167-2789}
\urldef\tempurl%
\url{https://doi.org/10.1016/0167-2789(88)90081-4}
\showDOI{\tempurl}


\bibitem[Hossin and M.N(2015)]%
        {metrics1}
\bibfield{author}{\bibinfo{person}{Mohammad Hossin} {and} \bibinfo{person}{Sulaiman M.N}.} \bibinfo{year}{2015}\natexlab{}.
\newblock \showarticletitle{A review on evaluation metrics for data classification evaluations}.
\newblock \bibinfo{journal}{\emph{International Journal of Data Mining \& Knowledge Management Process}} \bibinfo{volume}{5}, \bibinfo{number}{2} (\bibinfo{year}{2015}), \bibinfo{pages}{1--11}.
\newblock
\urldef\tempurl%
\url{https://doi.org/10.5121/ijdkp.2015.5201}
\showDOI{\tempurl}


\bibitem[Hsu et~al\mbox{.}(2020)]%
        {ECG}
\bibfield{author}{\bibinfo{person}{Yu-Liang Hsu}, \bibinfo{person}{Jeen-Shing Wang}, \bibinfo{person}{Wei-Chun Chiang}, {and} \bibinfo{person}{Chien-Han Hung}.} \bibinfo{year}{2020}\natexlab{}.
\newblock \showarticletitle{Automatic {ECG}-based emotion recognition in music listening}.
\newblock \bibinfo{journal}{\emph{IEEE Transactions on Affective Computing}} \bibinfo{volume}{11}, \bibinfo{number}{1} (\bibinfo{year}{2020}), \bibinfo{pages}{85--99}.
\newblock
\urldef\tempurl%
\url{https://doi.org/10.1109/TAFFC.2017.2781732}
\showDOI{\tempurl}


\bibitem[Hu et~al\mbox{.}(2022)]%
        {HKU956}
\bibfield{author}{\bibinfo{person}{Xiao Hu}, \bibinfo{person}{Fanjie Li}, {and} \bibinfo{person}{Ruilun Liu}.} \bibinfo{year}{2022}\natexlab{}.
\newblock \showarticletitle{Detecting music-induced emotion based on acoustic analysis and physiological sensing: A multimodal approach}.
\newblock \bibinfo{journal}{\emph{Applied Sciences}} \bibinfo{volume}{12}, \bibinfo{number}{18} (\bibinfo{year}{2022}), \bibinfo{pages}{9354}.
\newblock
\showISSN{2076-3417}
\urldef\tempurl%
\url{https://doi.org/10.3390/app12189354}
\showDOI{\tempurl}


\bibitem[Hu et~al\mbox{.}(2018)]%
        {psy_1}
\bibfield{author}{\bibinfo{person}{Xiao Hu}, \bibinfo{person}{Fanjie Li}, {and} \bibinfo{person}{Jeremy Ng}.} \bibinfo{year}{2018}\natexlab{}.
\newblock \showarticletitle{On the relationships between music-induced emotion and physiological signals}. In \bibinfo{booktitle}{\emph{International Society for Music Information Retrieval Conference (ISMIR)}}. \bibinfo{publisher}{International Society for Music Information Retrieval}, \bibinfo{address}{Paris, France}, \bibinfo{pages}{362--369}.
\newblock
\urldef\tempurl%
\url{https://archives.ismir.net/ismir2018/paper/000115.pdf}
\showURL{%
\tempurl}


\bibitem[Huang et~al\mbox{.}(2024)]%
        {AI_EMO_2}
\bibfield{author}{\bibinfo{person}{Jingyue Huang}, \bibinfo{person}{Ke Chen}, {and} \bibinfo{person}{Yi-Hsuan Yang}.} \bibinfo{year}{2024}\natexlab{}.
\newblock \bibinfo{title}{Emotion-driven Piano Music Generation via Two-stage Disentanglement and Functional Representation}.
\newblock
\newblock
\showeprint[arxiv]{2407.20955}~[cs.SD]
\urldef\tempurl%
\url{https://arxiv.org/abs/2407.20955}
\showURL{%
\tempurl}


\bibitem[Huang et~al\mbox{.}(2016)]%
        {Huang}
\bibfield{author}{\bibinfo{person}{Moyuan Huang}, \bibinfo{person}{Wenge Rong}, \bibinfo{person}{Tom Arjannikov}, \bibinfo{person}{Nan Jiang}, {and} \bibinfo{person}{Zhang Xiong}.} \bibinfo{year}{2016}\natexlab{}.
\newblock \showarticletitle{Bi-modal deep {Boltzmann} machine based musical emotion classification}. In \bibinfo{booktitle}{\emph{Artificial Neural Networks and Machine Learning (ICANN)}}. \bibinfo{publisher}{Springer International Publishing}, \bibinfo{address}{Cham, Switzerland}, \bibinfo{pages}{199--207}.
\newblock
\showISBNx{978-3-319-44781-0}
\urldef\tempurl%
\url{https://doi.org/10.1007/978-3-319-44781-0_24}
\showDOI{\tempurl}


\bibitem[Huertas-Garc{\'\i}a et~al\mbox{.}(2022)]%
        {huertas}
\bibfield{author}{\bibinfo{person}{{\'A}lvaro Huertas-Garc{\'\i}a}, \bibinfo{person}{Helena Liz}, \bibinfo{person}{Guillermo Villar-Rodr{\'\i}guez}, \bibinfo{person}{Alejandro Mart{\'\i}n}, \bibinfo{person}{Javier Huertas-Tato}, {and} \bibinfo{person}{David Camacho}.} \bibinfo{year}{2022}\natexlab{}.
\newblock \showarticletitle{{AIDA-UPM at SemEval-2022 Task 5}: Exploring multimodal late information fusion for multimedia automatic misogyny identification}. In \bibinfo{booktitle}{\emph{International Workshop on Semantic Evaluation (SemEval)}}. \bibinfo{publisher}{Association for Computational Linguistics}, \bibinfo{address}{Seattle, USA}, \bibinfo{pages}{771--779}.
\newblock
\urldef\tempurl%
\url{https://doi.org/10.18653/v1/2022.semeval-1.107}
\showDOI{\tempurl}


\bibitem[Hung et~al\mbox{.}(2021)]%
        {emopia}
\bibfield{author}{\bibinfo{person}{Hsiao-Tzu Hung}, \bibinfo{person}{Joann Ching}, \bibinfo{person}{Seungheon Doh}, \bibinfo{person}{Nabin Kim}, \bibinfo{person}{Juhan Nam}, {and} \bibinfo{person}{Yi-Hsuan Yang}.} \bibinfo{year}{2021}\natexlab{}.
\newblock \bibinfo{title}{EMOPIA: A multi-modal pop piano dataset for emotion recognition and emotion-based music generation}.
\newblock
\newblock
\showeprint[arxiv]{2108.01374}
\urldef\tempurl%
\url{https://arxiv.org/abs/2108.01374}
\showURL{%
\tempurl}


\bibitem[Huq et~al\mbox{.}(2010)]%
        {continuous_survey}
\bibfield{author}{\bibinfo{person}{Arefin Huq}, \bibinfo{person}{Juan~Pablo Bello}, {and} \bibinfo{person}{Robert Rowe}.} \bibinfo{year}{2010}\natexlab{}.
\newblock \showarticletitle{Automated music emotion recognition: A systematic evaluation}.
\newblock \bibinfo{journal}{\emph{Journal of New Music Research}} \bibinfo{volume}{39}, \bibinfo{number}{3} (\bibinfo{year}{2010}), \bibinfo{pages}{227--244}.
\newblock
\urldef\tempurl%
\url{https://doi.org/10.1080/09298215.2010.513733}
\showDOI{\tempurl}


\bibitem[Hussain et~al\mbox{.}(2024)]%
        {Hussain-2024}
\bibfield{author}{\bibinfo{person}{Mazhar Hussain}, \bibinfo{person}{Mattias O'Nils}, \bibinfo{person}{Jan Lundgren}, {and} \bibinfo{person}{Seyed~Jalaleddin Mousavirad}.} \bibinfo{year}{2024}\natexlab{}.
\newblock \showarticletitle{A comprehensive review on deep learning-based data fusion}.
\newblock \bibinfo{journal}{\emph{IEEE Access}}  \bibinfo{volume}{12} (\bibinfo{year}{2024}), \bibinfo{pages}{180093--180124}.
\newblock
\urldef\tempurl%
\url{https://doi.org/10.1109/ACCESS.2024.3508271}
\showDOI{\tempurl}


\bibitem[Iman et~al\mbox{.}(2023)]%
        {transfer_learning}
\bibfield{author}{\bibinfo{person}{Mohammadreza Iman}, \bibinfo{person}{Hamid~Reza Arabnia}, {and} \bibinfo{person}{Khaled Rasheed}.} \bibinfo{year}{2023}\natexlab{}.
\newblock \showarticletitle{A review of deep transfer learning and recent advancements}.
\newblock \bibinfo{journal}{\emph{Technologies}} \bibinfo{volume}{11}, \bibinfo{number}{2} (\bibinfo{year}{2023}), \bibinfo{pages}{40}.
\newblock
\showISSN{2227-7080}
\urldef\tempurl%
\url{https://doi.org/10.3390/technologies11020040}
\showDOI{\tempurl}


\bibitem[Jeon et~al\mbox{.}(2017)]%
        {2017-1}
\bibfield{author}{\bibinfo{person}{Byungsoo Jeon}, \bibinfo{person}{Chanju Kim}, \bibinfo{person}{Adrian Kim}, \bibinfo{person}{Dongwon Kim}, \bibinfo{person}{Jangyeon Park}, {and} \bibinfo{person}{Jung-Woo Ha}.} \bibinfo{year}{2017}\natexlab{}.
\newblock \showarticletitle{Music emotion recognition via end-to-end multimodal neural networks}. In \bibinfo{booktitle}{\emph{ACM Conference on Recommender Systems (RecSys)}}. \bibinfo{publisher}{ACM}, \bibinfo{address}{Como, Italy}, \bibinfo{pages}{1--2}.
\newblock
\urldef\tempurl%
\url{https://api.semanticscholar.org/CorpusID:416794}
\showURL{%
\tempurl}


\bibitem[Jeong and Lee(2016)]%
        {Jeong}
\bibfield{author}{\bibinfo{person}{Il-Young Jeong} {and} \bibinfo{person}{Kyogu Lee}.} \bibinfo{year}{2016}\natexlab{}.
\newblock \showarticletitle{Learning temporal features using a deep neural network and its application to music genre classification}. In \bibinfo{booktitle}{\emph{International Society for Music Information Retrieval Conference (ISMIR)}}. \bibinfo{publisher}{International Society for Music Information Retrieval}, \bibinfo{address}{New York, NY, USA}, \bibinfo{pages}{434--440}.
\newblock
\urldef\tempurl%
\url{https://archives.ismir.net/ismir2016/paper/000159.pdf}
\showURL{%
\tempurl}


\bibitem[Jitendra and Yalavarthi(2020)]%
        {audio_Feat}
\bibfield{author}{\bibinfo{person}{Mukkamala Jitendra} {and} \bibinfo{person}{Radhika Yalavarthi}.} \bibinfo{year}{2020}\natexlab{}.
\newblock \showarticletitle{A review: Music feature extraction from an audio signal}.
\newblock \bibinfo{journal}{\emph{International Journal of Advanced Trends in Computer Science and Engineering}}  \bibinfo{volume}{9} (\bibinfo{year}{2020}), \bibinfo{pages}{973--980}.
\newblock
\urldef\tempurl%
\url{https://doi.org/10.30534/ijatcse/2020/11922020}
\showDOI{\tempurl}


\bibitem[Joseph and Lekamge(2019)]%
        {icac_survey}
\bibfield{author}{\bibinfo{person}{Charles Joseph} {and} \bibinfo{person}{Sugeeswari Lekamge}.} \bibinfo{year}{2019}\natexlab{}.
\newblock \showarticletitle{Machine learning approaches for emotion classification of music: A systematic literature review}. In \bibinfo{booktitle}{\emph{International Conference on Advancements in Computing (ICAC)}}. \bibinfo{publisher}{IEEE}, \bibinfo{address}{Malabe, Sri Lanka}, \bibinfo{pages}{334--339}.
\newblock
\urldef\tempurl%
\url{https://doi.org/10.1109/ICAC49085.2019.9103378}
\showDOI{\tempurl}


\bibitem[Juslin(2001)]%
        {juslin}
\bibfield{author}{\bibinfo{person}{Patrik~N. Juslin}.} \bibinfo{year}{2001}\natexlab{}.
\newblock \showarticletitle{Communicating emotion in music performance: A review and theoretical framework}. In \bibinfo{booktitle}{\emph{Music and Emotion: Theory and Research}}. \bibinfo{publisher}{Oxford Academic}, \bibinfo{address}{New York, NY, USA}, \bibinfo{pages}{309--338}.
\newblock
\showISBNx{9780192631886}
\urldef\tempurl%
\url{https://doi.org/10.1093/oso/9780192631886.003.0014}
\showDOI{\tempurl}


\bibitem[Juslin and Laukka(2004)]%
        {Patrik}
\bibfield{author}{\bibinfo{person}{Patrik~N. Juslin} {and} \bibinfo{person}{Petri Laukka}.} \bibinfo{year}{2004}\natexlab{}.
\newblock \showarticletitle{Expression, perception, and induction of musical emotions: A review and a questionnaire study of everyday listening}.
\newblock \bibinfo{journal}{\emph{Journal of New Music Research}} \bibinfo{volume}{33}, \bibinfo{number}{3} (\bibinfo{year}{2004}), \bibinfo{pages}{217--238}.
\newblock
\urldef\tempurl%
\url{https://doi.org/10.1080/0929821042000317813}
\showDOI{\tempurl}


\bibitem[Juslin et~al\mbox{.}(2022)]%
        {background}
\bibfield{author}{\bibinfo{person}{Patrik~N. Juslin}, \bibinfo{person}{Laura~S. Sakka}, \bibinfo{person}{Gonçalo~T. Barradas}, {and} \bibinfo{person}{Olivier Lartillot}.} \bibinfo{year}{2022}\natexlab{}.
\newblock \showarticletitle{Emotions, mechanisms, and individual differences in music listening: A stratified random sampling approach}.
\newblock \bibinfo{journal}{\emph{Music Perception}} \bibinfo{volume}{40}, \bibinfo{number}{1} (\bibinfo{year}{2022}), \bibinfo{pages}{55--86}.
\newblock
\showISSN{0730-7829}
\urldef\tempurl%
\url{https://doi.org/10.1525/mp.2022.40.1.55}
\showDOI{\tempurl}


\bibitem[Juslin and Västfjäll(2008)]%
        {mood_and_em}
\bibfield{author}{\bibinfo{person}{Patrik~N. Juslin} {and} \bibinfo{person}{Daniel Västfjäll}.} \bibinfo{year}{2008}\natexlab{}.
\newblock \showarticletitle{Emotional responses to music: The need to consider underlying mechanisms}.
\newblock \bibinfo{journal}{\emph{Behavioral and Brain Sciences}} \bibinfo{volume}{31}, \bibinfo{number}{5} (\bibinfo{year}{2008}), \bibinfo{pages}{559–575}.
\newblock
\urldef\tempurl%
\url{https://doi.org/10.1017/S0140525X08005293}
\showDOI{\tempurl}


\bibitem[Kang et~al\mbox{.}(2023)]%
        {AI_EMO}
\bibfield{author}{\bibinfo{person}{Chenfei Kang}, \bibinfo{person}{Peiling Lu}, \bibinfo{person}{Botao Yu}, \bibinfo{person}{Xu Tan}, \bibinfo{person}{Wei Ye}, \bibinfo{person}{Shikun Zhang}, {and} \bibinfo{person}{Jiang Bian}.} \bibinfo{year}{2023}\natexlab{}.
\newblock \bibinfo{title}{EmoGen: Eliminating Subjective Bias in Emotional Music Generation}.
\newblock
\newblock
\showeprint[arxiv]{2307.01229}~[cs.SD]
\urldef\tempurl%
\url{https://arxiv.org/abs/2307.01229}
\showURL{%
\tempurl}


\bibitem[Kemper and Danhauer(2005)]%
        {therapy}
\bibfield{author}{\bibinfo{person}{Kathi~J Kemper} {and} \bibinfo{person}{Suzanne~C Danhauer}.} \bibinfo{year}{2005}\natexlab{}.
\newblock \showarticletitle{Music as therapy}.
\newblock \bibinfo{journal}{\emph{Southern Medical Journal}} \bibinfo{volume}{98}, \bibinfo{number}{3} (\bibinfo{year}{2005}), \bibinfo{pages}{282--288}.
\newblock
\urldef\tempurl%
\url{https://doi.org/10.1097/01.smj.0000154773.11986.39}
\showDOI{\tempurl}


\bibitem[Kim et~al\mbox{.}(2010)]%
        {ISMIR-2010}
\bibfield{author}{\bibinfo{person}{Youngmoo Kim}, \bibinfo{person}{Erik Schmidt}, \bibinfo{person}{Raymond Migneco}, \bibinfo{person}{Brandon Morton}, \bibinfo{person}{Patrick Richardson}, \bibinfo{person}{Jeffrey Scott}, \bibinfo{person}{Jacquelin Speck}, {and} \bibinfo{person}{Douglas Turnbull}.} \bibinfo{year}{2010}\natexlab{}.
\newblock \showarticletitle{Music emotion recognition: A state of the art review}. In \bibinfo{booktitle}{\emph{International Society for Music Information Retrieval Conference (ISMIR)}}. \bibinfo{publisher}{International Society for Music Information Retrieval}, \bibinfo{address}{Utrecht, The Netherlands}, \bibinfo{pages}{255--266}.
\newblock
\urldef\tempurl%
\url{https://archives.ismir.net/ismir2010/paper/000045.pdf}
\showURL{%
\tempurl}


\bibitem[Kittler et~al\mbox{.}(1998)]%
        {LPF}
\bibfield{author}{\bibinfo{person}{Josef Kittler}, \bibinfo{person}{Mr. Hatef}, \bibinfo{person}{Robert Duin}, {and} \bibinfo{person}{Jiri Matas}.} \bibinfo{year}{1998}\natexlab{}.
\newblock \showarticletitle{On combining classifiers}.
\newblock \bibinfo{journal}{\emph{IEEE Transactions on Pattern Analysis and Machine Intelligence}} \bibinfo{volume}{20}, \bibinfo{number}{3} (\bibinfo{year}{1998}), \bibinfo{pages}{226--239}.
\newblock
\urldef\tempurl%
\url{https://doi.org/10.1109/34.667881}
\showDOI{\tempurl}


\bibitem[Koelstra et~al\mbox{.}(2012)]%
        {deap}
\bibfield{author}{\bibinfo{person}{Sander Koelstra}, \bibinfo{person}{Christian Muhl}, \bibinfo{person}{Mohammad Soleymani}, \bibinfo{person}{Jong-Seok Lee}, \bibinfo{person}{Ashkan Yazdani}, \bibinfo{person}{Touradj Ebrahimi}, \bibinfo{person}{Thierry Pun}, \bibinfo{person}{Anton Nijholt}, {and} \bibinfo{person}{Ioannis Patras}.} \bibinfo{year}{2012}\natexlab{}.
\newblock \showarticletitle{{DEAP}: A database for emotion analysis using physiological signals}.
\newblock \bibinfo{journal}{\emph{IEEE Transactions on Affective Computing}} \bibinfo{volume}{3}, \bibinfo{number}{1} (\bibinfo{year}{2012}), \bibinfo{pages}{18--31}.
\newblock
\urldef\tempurl%
\url{https://doi.org/10.1109/T-AFFC.2011.15}
\showDOI{\tempurl}


\bibitem[Koh et~al\mbox{.}(2023)]%
        {MERP}
\bibfield{author}{\bibinfo{person}{En~Yan Koh}, \bibinfo{person}{Kin~Wai Cheuk}, \bibinfo{person}{Kwan~Yee Heung}, \bibinfo{person}{Kat~R. Agres}, {and} \bibinfo{person}{Dorien Herremans}.} \bibinfo{year}{2023}\natexlab{}.
\newblock \showarticletitle{{MERP}: A music dataset with emotion ratings and raters’ profile information}.
\newblock \bibinfo{journal}{\emph{Sensors}} \bibinfo{volume}{23}, \bibinfo{number}{1} (\bibinfo{year}{2023}), \bibinfo{pages}{382}.
\newblock
\showISSN{1424-8220}
\urldef\tempurl%
\url{https://doi.org/10.3390/s23010382}
\showDOI{\tempurl}


\bibitem[Lartillot et~al\mbox{.}(2007)]%
        {MIRtoolbox}
\bibfield{author}{\bibinfo{person}{Olivier Lartillot}, \bibinfo{person}{Petri Toiviainen}, {and} \bibinfo{person}{Tuomas Eerola}.} \bibinfo{year}{2007}\natexlab{}.
\newblock \showarticletitle{A {Matlab} toolbox for music information retrieval}. In \bibinfo{booktitle}{\emph{Annual Conference of the Gesellschaft f{\"u}r Klassifikation}}. \bibinfo{publisher}{Springer}, \bibinfo{address}{Berlin, Heidelberg}, \bibinfo{pages}{261--268}.
\newblock
\urldef\tempurl%
\url{https://api.semanticscholar.org/CorpusID:17342536}
\showURL{%
\tempurl}


\bibitem[Laurier et~al\mbox{.}(2008)]%
        {2008_1}
\bibfield{author}{\bibinfo{person}{Cyril Laurier}, \bibinfo{person}{Jens Grivolla}, {and} \bibinfo{person}{Perfecto Herrera}.} \bibinfo{year}{2008}\natexlab{}.
\newblock \showarticletitle{Multimodal music mood classification using audio and lyrics}. In \bibinfo{booktitle}{\emph{International Conference on Machine Learning and Applications (ICMLA)}}. \bibinfo{publisher}{IEEE}, \bibinfo{address}{San Diego, CA, USA}, \bibinfo{pages}{688--693}.
\newblock
\urldef\tempurl%
\url{https://doi.org/10.1109/ICMLA.2008.96}
\showDOI{\tempurl}


\bibitem[Lazarus(1995)]%
        {Lazarus}
\bibfield{author}{\bibinfo{person}{Richard~S. Lazarus}.} \bibinfo{year}{1995}\natexlab{}.
\newblock \showarticletitle{Vexing research problems inherent in cognitive-mediational theories of emotion- and some solutions}.
\newblock \bibinfo{journal}{\emph{Psychological Inquiry}} \bibinfo{volume}{6}, \bibinfo{number}{3} (\bibinfo{year}{1995}), \bibinfo{pages}{183--196}.
\newblock
\urldef\tempurl%
\url{https://doi.org/10.1207/s15327965pli0603\_1}
\showDOI{\tempurl}


\bibitem[Lindquist and Barrett(2008)]%
        {Lindquist}
\bibfield{author}{\bibinfo{person}{Kristen~A. Lindquist} {and} \bibinfo{person}{Lisa~Feldman Barrett}.} \bibinfo{year}{2008}\natexlab{}.
\newblock \showarticletitle{Constructing emotion: The experience of fear as a conceptual act}.
\newblock \bibinfo{journal}{\emph{Psychological Science}} \bibinfo{volume}{19}, \bibinfo{number}{9} (\bibinfo{year}{2008}), \bibinfo{pages}{898--903}.
\newblock
\urldef\tempurl%
\url{https://doi.org/10.1111/j.1467-9280.2008.02174.x}
\showDOI{\tempurl}


\bibitem[Liu and Tan(2020)]%
        {paper08}
\bibfield{author}{\bibinfo{person}{Gaojun Liu} {and} \bibinfo{person}{Zhiyuan Tan}.} \bibinfo{year}{2020}\natexlab{}.
\newblock \showarticletitle{Research on multi-modal music emotion classification based on audio and lyrics}. In \bibinfo{booktitle}{\emph{IEEE Information Technology, Networking, Electronic and Automation Control Conference (ITNEC)}}. \bibinfo{publisher}{IEEE}, \bibinfo{address}{Chongqing, China}, \bibinfo{pages}{2331--2335}.
\newblock
\urldef\tempurl%
\url{https://doi.org/10.1109/ITNEC48623.2020.9084846}
\showDOI{\tempurl}


\bibitem[Liu et~al\mbox{.}(2023)]%
        {paper20}
\bibfield{author}{\bibinfo{person}{Zhiyuan Liu}, \bibinfo{person}{Wei Xu}, \bibinfo{person}{Wenping Zhang}, {and} \bibinfo{person}{Qiqi Jiang}.} \bibinfo{year}{2023}\natexlab{}.
\newblock \showarticletitle{An emotion-based personalized music recommendation framework for emotion improvement}.
\newblock \bibinfo{journal}{\emph{Information Processing \& Management}} \bibinfo{volume}{60}, \bibinfo{number}{3} (\bibinfo{year}{2023}), \bibinfo{pages}{103256}.
\newblock
\showISSN{0306-4573}
\urldef\tempurl%
\url{https://doi.org/10.1016/j.ipm.2022.103256}
\showDOI{\tempurl}


\bibitem[Logan(2000)]%
        {logan}
\bibfield{author}{\bibinfo{person}{Beth Logan}.} \bibinfo{year}{2000}\natexlab{}.
\newblock \showarticletitle{Mel frequency cepstral coefficients for music modeling}. In \bibinfo{booktitle}{\emph{International Symposium on Music Information Retrieval (MUSIC IR)}}. \bibinfo{publisher}{University of Massachusetts}, \bibinfo{address}{Plymouth, Massachusetts, USA}, \bibinfo{pages}{1--2}.
\newblock
\urldef\tempurl%
\url{https://ismir2000.ismir.net/papers/logan_abs.pdf}
\showURL{%
\tempurl}


\bibitem[Louro et~al\mbox{.}(2024)]%
        {merge}
\bibfield{author}{\bibinfo{person}{Pedro~Lima Louro}, \bibinfo{person}{Hugo Redinho}, \bibinfo{person}{Ricardo Santos}, \bibinfo{person}{Ricardo Malheiro}, \bibinfo{person}{Renato Panda}, {and} \bibinfo{person}{Rui~Pedro Paiva}.} \bibinfo{year}{2024}\natexlab{}.
\newblock \bibinfo{title}{MERGE -- A bimodal dataset for static music emotion recognition}.
\newblock
\newblock
\showeprint[arxiv]{2407.06060}
\urldef\tempurl%
\url{https://arxiv.org/abs/2407.06060}
\showURL{%
\tempurl}


\bibitem[Malheiro et~al\mbox{.}(2013)]%
        {malheiro}
\bibfield{author}{\bibinfo{person}{Ricardo Malheiro}, \bibinfo{person}{Renato Panda}, \bibinfo{person}{Paulo Gomes}, {and} \bibinfo{person}{Rui~Pedro Paiva}.} \bibinfo{year}{2013}\natexlab{}.
\newblock \showarticletitle{Music emotion recognition from lyrics: A comparative study}. In \bibinfo{booktitle}{\emph{International Workshop on Music and Machine Learning (MML)}}. \bibinfo{publisher}{MML}, \bibinfo{address}{Prague, Czech Republic}, \bibinfo{pages}{1--4}.
\newblock
\urldef\tempurl%
\url{https://hdl.handle.net/10316/95165}
\showURL{%
\tempurl}


\bibitem[Mohammad and Turney(2013)]%
        {Mohammad}
\bibfield{author}{\bibinfo{person}{Saif~M. Mohammad} {and} \bibinfo{person}{Peter~D. Turney}.} \bibinfo{year}{2013}\natexlab{}.
\newblock \showarticletitle{Crowdsourcing a word–emotion association lexicon}.
\newblock \bibinfo{journal}{\emph{Computational Intelligence}} \bibinfo{volume}{29}, \bibinfo{number}{3} (\bibinfo{year}{2013}), \bibinfo{pages}{436--465}.
\newblock
\urldef\tempurl%
\url{https://doi.org/10.1111/j.1467-8640.2012.00460.x}
\showDOI{\tempurl}


\bibitem[Nagaraju and Sadanandam(2024)]%
        {optic_flow_motion}
\bibfield{author}{\bibinfo{person}{Pampati Nagaraju} {and} \bibinfo{person}{Manchala Sadanandam}.} \bibinfo{year}{2024}\natexlab{}.
\newblock \showarticletitle{Advancements in motion detection within video streams through the integration of optical flow estimation and {3D}-convolutional neural network architectures}.
\newblock \bibinfo{journal}{\emph{Journal of Electrical Systems}} \bibinfo{volume}{20}, \bibinfo{number}{6} (\bibinfo{year}{2024}), \bibinfo{pages}{2502--2517}.
\newblock
\urldef\tempurl%
\url{https://doi.org/10.52783/jes.3238}
\showDOI{\tempurl}


\bibitem[Naji et~al\mbox{.}(2013)]%
        {ECG2}
\bibfield{author}{\bibinfo{person}{Mohsen Naji}, \bibinfo{person}{Mohammad Firoozabadi}, {and} \bibinfo{person}{Parviz Azadfallah}.} \bibinfo{year}{2013}\natexlab{}.
\newblock \showarticletitle{Classification of music-induced emotions based on information fusion of forehead biosignals and electrocardiogram}.
\newblock \bibinfo{journal}{\emph{Cognitive Computation}}  \bibinfo{volume}{6} (\bibinfo{year}{2013}), \bibinfo{pages}{241--252}.
\newblock
\urldef\tempurl%
\url{https://doi.org/10.1007/s12559-013-9239-7}
\showDOI{\tempurl}


\bibitem[Panda et~al\mbox{.}(2018)]%
        {4Q}
\bibfield{author}{\bibinfo{person}{Renato Panda}, \bibinfo{person}{Ricardo Malheiro}, {and} \bibinfo{person}{Rui~Pedro Paiva}.} \bibinfo{year}{2018}\natexlab{}.
\newblock \showarticletitle{Musical texture and expressivity features for music emotion recognition}. In \bibinfo{booktitle}{\emph{International Society for Music Information Retrieval Conference (ISMIR)}}. \bibinfo{publisher}{International Society for Music Information Retrieval}, \bibinfo{address}{Paris, France}, \bibinfo{pages}{383--391}.
\newblock
\urldef\tempurl%
\url{https://api.semanticscholar.org/CorpusID:53875359}
\showURL{%
\tempurl}


\bibitem[Panda et~al\mbox{.}(2013)]%
        {mirex-like}
\bibfield{author}{\bibinfo{person}{Renato Panda}, \bibinfo{person}{Ricardo Malheiro}, \bibinfo{person}{Bruno Rocha}, \bibinfo{person}{António Oliveira}, {and} \bibinfo{person}{Rui~Pedro Paiva}.} \bibinfo{year}{2013}\natexlab{}.
\newblock \showarticletitle{Multi-modal music emotion recognition: A new dataset, methodology and comparative analysis}. In \bibinfo{booktitle}{\emph{International Symposium on Computer Music Multidisciplinary Research (CMMR)}}. \bibinfo{publisher}{Springer Verlag}, \bibinfo{address}{Marseille, France}, \bibinfo{pages}{570--582}.
\newblock
\urldef\tempurl%
\url{https://hdl.handle.net/10316/94095}
\showURL{%
\tempurl}


\bibitem[Pandeya et~al\mbox{.}(2021)]%
        {paper02}
\bibfield{author}{\bibinfo{person}{Yagya~Raj Pandeya}, \bibinfo{person}{Bhuwan Bhattarai}, {and} \bibinfo{person}{Joonwhoan Lee}.} \bibinfo{year}{2021}\natexlab{}.
\newblock \showarticletitle{Deep-learning-based multimodal emotion classification for music videos}.
\newblock \bibinfo{journal}{\emph{Sensors}} \bibinfo{volume}{21}, \bibinfo{number}{14} (\bibinfo{year}{2021}), \bibinfo{pages}{4927}.
\newblock
\showISSN{1424-8220}
\urldef\tempurl%
\url{https://doi.org/10.3390/s21144927}
\showDOI{\tempurl}


\bibitem[Pegoraro~Santana et~al\mbox{.}(2020)]%
        {music4all}
\bibfield{author}{\bibinfo{person}{Igor~André Pegoraro~Santana}, \bibinfo{person}{Fabio Pinhelli}, \bibinfo{person}{Juliano Donini}, \bibinfo{person}{Leonardo Catharin}, \bibinfo{person}{Rafael~Biazus Mangolin}, \bibinfo{person}{Yandre Maldonado e~Gomes da Costa}, \bibinfo{person}{Valéria Delisandra~Feltrim}, {and} \bibinfo{person}{Marcos~Aurélio Domingues}.} \bibinfo{year}{2020}\natexlab{}.
\newblock \showarticletitle{{Music4All}: A new music database and its applications}. In \bibinfo{booktitle}{\emph{2020 International Conference on Systems, Signals and Image Processing (IWSSIP)}}. \bibinfo{publisher}{IEEE}, \bibinfo{address}{Niteroi, Brazil}, \bibinfo{pages}{399--404}.
\newblock
\urldef\tempurl%
\url{https://doi.org/10.1109/IWSSIP48289.2020.9145170}
\showDOI{\tempurl}


\bibitem[Plutchik and Kellerman(2013)]%
        {plutchikBook}
\bibfield{author}{\bibinfo{person}{Robert Plutchik} {and} \bibinfo{person}{Henry Kellerman}.} \bibinfo{year}{2013}\natexlab{}.
\newblock \bibinfo{booktitle}{\emph{EMOTION: Theory, Research, and Experience}}. \bibinfo{series}{Theories of emotion}, Vol.~\bibinfo{volume}{1}.
\newblock \bibinfo{publisher}{Academic press}, \bibinfo{address}{New York, NY, USA}.
\newblock
\urldef\tempurl%
\url{https://www.sciencedirect.com/book/9780125587013/}
\showURL{%
\tempurl}


\bibitem[Posner et~al\mbox{.}(2005)]%
        {Henver}
\bibfield{author}{\bibinfo{person}{Jonathan Posner}, \bibinfo{person}{James~A Russell}, {and} \bibinfo{person}{Bradley~S Peterson}.} \bibinfo{year}{2005}\natexlab{}.
\newblock \showarticletitle{The circumplex model of affect: An integrative approach to affective neuroscience, cognitive development, and psychopathology}.
\newblock \bibinfo{journal}{\emph{Development and Psychopathology}} \bibinfo{volume}{17}, \bibinfo{number}{3} (\bibinfo{year}{2005}), \bibinfo{pages}{715--734}.
\newblock
\urldef\tempurl%
\url{https://doi.org/10.1017/S0954579405050340}
\showDOI{\tempurl}


\bibitem[Pyrovolakis et~al\mbox{.}(2022)]%
        {paper01}
\bibfield{author}{\bibinfo{person}{Konstantinos Pyrovolakis}, \bibinfo{person}{Paraskevi Tzouveli}, {and} \bibinfo{person}{Giorgos Stamou}.} \bibinfo{year}{2022}\natexlab{}.
\newblock \showarticletitle{Multi-modal song mood detection with deep learning}.
\newblock \bibinfo{journal}{\emph{Sensors}} \bibinfo{volume}{22}, \bibinfo{number}{3} (\bibinfo{year}{2022}), \bibinfo{pages}{1065}.
\newblock
\showISSN{1424-8220}
\urldef\tempurl%
\url{https://doi.org/10.3390/s22031065}
\showDOI{\tempurl}


\bibitem[Ramachandram and Taylor(2017)]%
        {Ramachandram-2017}
\bibfield{author}{\bibinfo{person}{Dhanesh Ramachandram} {and} \bibinfo{person}{Graham~W. Taylor}.} \bibinfo{year}{2017}\natexlab{}.
\newblock \showarticletitle{Deep multimodal learning: A survey on recent advances and trends}.
\newblock \bibinfo{journal}{\emph{IEEE Signal Processing Magazine}} \bibinfo{volume}{34}, \bibinfo{number}{6} (\bibinfo{year}{2017}), \bibinfo{pages}{96--108}.
\newblock
\urldef\tempurl%
\url{https://doi.org/10.1109/MSP.2017.2738401}
\showDOI{\tempurl}


\bibitem[Russell(1980)]%
        {russell}
\bibfield{author}{\bibinfo{person}{James Russell}.} \bibinfo{year}{1980}\natexlab{}.
\newblock \showarticletitle{A circumplex model of affect}.
\newblock \bibinfo{journal}{\emph{Journal of Personality and Social Psychology}} \bibinfo{volume}{39}, \bibinfo{number}{6} (\bibinfo{year}{1980}), \bibinfo{pages}{1161--1178}.
\newblock
\urldef\tempurl%
\url{https://doi.org/10.1037/h0077714}
\showDOI{\tempurl}


\bibitem[S. and Rajeev(2023)]%
        {Sujeesha}
\bibfield{author}{\bibinfo{person}{Sujeesha~A. S.} {and} \bibinfo{person}{Rajan Rajeev}.} \bibinfo{year}{2023}\natexlab{}.
\newblock \showarticletitle{Transformer-based automatic music mood classification using multi-modal framework}.
\newblock \bibinfo{journal}{\emph{Journal of Computer Science and Technology}} \bibinfo{volume}{23}, \bibinfo{number}{1} (\bibinfo{year}{2023}), \bibinfo{pages}{e02}.
\newblock
\urldef\tempurl%
\url{https://doi.org/10.24215/16666038.23.e02}
\showDOI{\tempurl}


\bibitem[Sams and Zahra(2023)]%
        {paper06}
\bibfield{author}{\bibinfo{person}{Andrew Sams} {and} \bibinfo{person}{Amalia Zahra}.} \bibinfo{year}{2023}\natexlab{}.
\newblock \showarticletitle{Multimodal music emotion recognition in {Indonesian} songs based on {CNN-LSTM, XLNet} transformers}.
\newblock \bibinfo{journal}{\emph{Bulletin of Electrical Engineering and Informatics}} \bibinfo{volume}{12}, \bibinfo{number}{1} (\bibinfo{year}{2023}), \bibinfo{pages}{355--364}.
\newblock
\urldef\tempurl%
\url{https://doi.org/10.11591/eei.v12i1.4231}
\showDOI{\tempurl}


\bibitem[Scherer(2001)]%
        {schrer}
\bibfield{author}{\bibinfo{person}{Klaus Scherer}.} \bibinfo{year}{2001}\natexlab{}.
\newblock \showarticletitle{Appraisal considered as a process of multilevel sequential checking}. In \bibinfo{booktitle}{\emph{Appraisal Processes in Emotion: Theory, Methods, Research}}. \bibinfo{publisher}{Oxford University Press}, \bibinfo{address}{Oxford, UK}, \bibinfo{pages}{92--120}.
\newblock
\showISBNx{9780195130072}
\urldef\tempurl%
\url{https://doi.org/10.1093/oso/9780195130072.003.0005}
\showDOI{\tempurl}


\bibitem[Scherer(2005)]%
        {GEW}
\bibfield{author}{\bibinfo{person}{Klaus~R. Scherer}.} \bibinfo{year}{2005}\natexlab{}.
\newblock \showarticletitle{What are emotions? {And} how can they be measured?}
\newblock \bibinfo{journal}{\emph{Social Science Information}} \bibinfo{volume}{44}, \bibinfo{number}{4} (\bibinfo{year}{2005}), \bibinfo{pages}{695--729}.
\newblock
\urldef\tempurl%
\url{https://doi.org/10.1177/0539018405058216}
\showDOI{\tempurl}


\bibitem[Sokolova and Lapalme(2009)]%
        {metrics2}
\bibfield{author}{\bibinfo{person}{Marina Sokolova} {and} \bibinfo{person}{Guy Lapalme}.} \bibinfo{year}{2009}\natexlab{}.
\newblock \showarticletitle{A systematic analysis of performance measures for classification tasks}.
\newblock \bibinfo{journal}{\emph{Information Processing \& Management}} \bibinfo{volume}{45}, \bibinfo{number}{4} (\bibinfo{year}{2009}), \bibinfo{pages}{427--437}.
\newblock
\showISSN{0306-4573}
\urldef\tempurl%
\url{https://doi.org/10.1016/j.ipm.2009.03.002}
\showDOI{\tempurl}


\bibitem[Soleymani et~al\mbox{.}(2013)]%
        {MediaEval}
\bibfield{author}{\bibinfo{person}{Mohammad Soleymani}, \bibinfo{person}{Micheal~N. Caro}, \bibinfo{person}{Erik~M. Schmidt}, \bibinfo{person}{Cheng-Ya Sha}, {and} \bibinfo{person}{Yi-Hsuan Yang}.} \bibinfo{year}{2013}\natexlab{}.
\newblock \showarticletitle{1000 songs for emotional analysis of music}. In \bibinfo{booktitle}{\emph{ACM International Workshop on Crowdsourcing for Multimedia (CrowdMM)}} (Barcelona, Spain) \emph{(\bibinfo{series}{CrowdMM '13})}. \bibinfo{publisher}{Association for Computing Machinery}, \bibinfo{address}{New York, NY, USA}, \bibinfo{pages}{1–6}.
\newblock
\showISBNx{9781450323963}
\urldef\tempurl%
\url{https://doi.org/10.1145/2506364.2506365}
\showDOI{\tempurl}


\bibitem[Strauß et~al\mbox{.}(2024)]%
        {EMMA}
\bibfield{author}{\bibinfo{person}{Hannah Strauß}, \bibinfo{person}{Julia Vigl}, \bibinfo{person}{Peer-Ole Jacobsen}, \bibinfo{person}{Martin Bayer}, \bibinfo{person}{Francesca Talamini}, \bibinfo{person}{Wolfgang Vigl}, \bibinfo{person}{Eva Zangerle}, {and} \bibinfo{person}{Marcel Zentner}.} \bibinfo{year}{2024}\natexlab{}.
\newblock \showarticletitle{{The Emotion-to-Music Mapping Atlas (EMMA)}: A systematically organized online database of emotionally evocative music excerpts}.
\newblock \bibinfo{journal}{\emph{Behavior Research Methods}}  \bibinfo{volume}{56} (\bibinfo{year}{2024}), \bibinfo{pages}{3560–3577}.
\newblock
\urldef\tempurl%
\url{https://doi.org/10.3758/s13428-024-02336-0}
\showDOI{\tempurl}


\bibitem[Sze et~al\mbox{.}(2017)]%
        {ext_2}
\bibfield{author}{\bibinfo{person}{Vivienne Sze}, \bibinfo{person}{Yu-Hsin Chen}, \bibinfo{person}{Tien-Ju Yang}, {and} \bibinfo{person}{Joel Emer}.} \bibinfo{year}{2017}\natexlab{}.
\newblock \bibinfo{title}{Efficient Processing of Deep Neural Networks: A Tutorial and Survey}.
\newblock
\newblock
\showeprint[arxiv]{1703.09039}~[cs.CV]
\urldef\tempurl%
\url{https://arxiv.org/abs/1703.09039}
\showURL{%
\tempurl}


\bibitem[Tatachar(2021)]%
        {metrics_reg}
\bibfield{author}{\bibinfo{person}{Abhishek~V. Tatachar}.} \bibinfo{year}{2021}\natexlab{}.
\newblock \showarticletitle{Comparative assessment of regression models based on model evaluation metrics}.
\newblock \bibinfo{journal}{\emph{International Journal of Innovative Technology and Exploring Engineering}} \bibinfo{volume}{8}, \bibinfo{number}{9} (\bibinfo{year}{2021}), \bibinfo{pages}{853--860}.
\newblock
\urldef\tempurl%
\url{https://www.irjet.net/archives/V8/i9/IRJET-V8I9127.pdf}
\showURL{%
\tempurl}


\bibitem[Taylor et~al\mbox{.}(2024)]%
        {ext_3}
\bibfield{author}{\bibinfo{person}{Jing~Wen Taylor}, \bibinfo{person}{Chuan Ching-Hua}, \bibinfo{person}{Anghelcev George}, \bibinfo{person}{Sar Sela}, \bibinfo{person}{T.~Yun Joseph}, {and} \bibinfo{person}{Xu Yanzhen}.} \bibinfo{year}{2024}\natexlab{}.
\newblock \showarticletitle{Infusing affective computing models into advertising research on emotions}.
\newblock \bibinfo{journal}{\emph{Journal of Advertising}} \bibinfo{volume}{53}, \bibinfo{number}{5} (\bibinfo{year}{2024}), \bibinfo{pages}{710--731}.
\newblock
\urldef\tempurl%
\url{https://doi.org/10.1080/00913367.2024.2409254}
\showDOI{\tempurl}
\showeprint{https://doi.org/10.1080/00913367.2024.2409254}


\bibitem[Thammasan et~al\mbox{.}(2017)]%
        {2017-2}
\bibfield{author}{\bibinfo{person}{Nattapong Thammasan}, \bibinfo{person}{Ken-ichi Fukui}, {and} \bibinfo{person}{Masayuki Numao}.} \bibinfo{year}{2017}\natexlab{}.
\newblock \showarticletitle{Multimodal fusion of {EEG} and musical features in music-emotion recognition}. In \bibinfo{booktitle}{\emph{AAAI Conference on Artificial Intelligence}}. \bibinfo{publisher}{AAAI Press}, \bibinfo{address}{San Francisco, California, USA}, \bibinfo{pages}{4991–4992}.
\newblock
\urldef\tempurl%
\url{https://doi.org/10.1609/aaai.v31i1.11112}
\showDOI{\tempurl}


\bibitem[Thao et~al\mbox{.}(2023)]%
        {emoMV}
\bibfield{author}{\bibinfo{person}{Ha~Thi~Phuong Thao}, \bibinfo{person}{Gemma Roig}, {and} \bibinfo{person}{Dorien Herremans}.} \bibinfo{year}{2023}\natexlab{}.
\newblock \showarticletitle{{EmoMV}: Affective music-video correspondence learning datasets for classification and retrieval}.
\newblock \bibinfo{journal}{\emph{Information Fusion}}  \bibinfo{volume}{91} (\bibinfo{year}{2023}), \bibinfo{pages}{64--79}.
\newblock
\showISSN{1566-2535}
\urldef\tempurl%
\url{https://doi.org/10.1016/j.inffus.2022.10.002}
\showDOI{\tempurl}


\bibitem[Thayer(1989)]%
        {thayers}
\bibfield{author}{\bibinfo{person}{Robert Thayer}.} \bibinfo{year}{1989}\natexlab{}.
\newblock \bibinfo{booktitle}{\emph{The Biopsychology of Mood and Arousal}}.
\newblock \bibinfo{publisher}{Oxford Academic}, \bibinfo{address}{New York, NY, USA}.
\newblock
\showISBNx{9780195068276}
\urldef\tempurl%
\url{https://doi.org/10.1093/oso/9780195068276.001.0001}
\showDOI{\tempurl}


\bibitem[Thompson(2013)]%
        {intervals}
\bibfield{author}{\bibinfo{person}{William~F. Thompson}.} \bibinfo{year}{2013}\natexlab{}.
\newblock \showarticletitle{Intervals and scales}. In \bibinfo{booktitle}{\emph{The Psychology of Music} (\bibinfo{edition}{3rd} ed.)}. \bibinfo{publisher}{Academic Press}, \bibinfo{address}{London, UK}, \bibinfo{pages}{107--140}.
\newblock
\urldef\tempurl%
\url{https://doi.org/10.1016/B978-0-12-381460-9.00004-3}
\showDOI{\tempurl}


\bibitem[Thompson et~al\mbox{.}(2001)]%
        {mozart}
\bibfield{author}{\bibinfo{person}{William~F. Thompson}, \bibinfo{person}{E.~Glenn Schellenberg}, {and} \bibinfo{person}{Gabriela Husain}.} \bibinfo{year}{2001}\natexlab{}.
\newblock \showarticletitle{Arousal, mood, and the {Mozart} effect}.
\newblock \bibinfo{journal}{\emph{Psychological Science}} \bibinfo{volume}{12}, \bibinfo{number}{3} (\bibinfo{year}{2001}), \bibinfo{pages}{248--251}.
\newblock
\urldef\tempurl%
\url{https://doi.org/10.1111/1467-9280.00345}
\showDOI{\tempurl}


\bibitem[Tong(2022)]%
        {paper09}
\bibfield{author}{\bibinfo{person}{Guiying Tong}.} \bibinfo{year}{2022}\natexlab{}.
\newblock \showarticletitle{Multimodal music emotion recognition method based on the combination of knowledge distillation and transfer learning}.
\newblock \bibinfo{journal}{\emph{Scientific Programming}} \bibinfo{volume}{2022}, \bibinfo{number}{1} (\bibinfo{year}{2022}), \bibinfo{pages}{2802573}.
\newblock
\urldef\tempurl%
\url{https://doi.org/10.1155/2022/2802573}
\showDOI{\tempurl}


\bibitem[Turnbull et~al\mbox{.}(2008)]%
        {CAL500}
\bibfield{author}{\bibinfo{person}{Douglas Turnbull}, \bibinfo{person}{Luke Barrington}, \bibinfo{person}{David Torres}, {and} \bibinfo{person}{Gert Lanckriet}.} \bibinfo{year}{2008}\natexlab{}.
\newblock \showarticletitle{Semantic annotation and retrieval of music and sound effects}.
\newblock \bibinfo{journal}{\emph{IEEE Transactions on Audio, Speech, and Language Processing}} \bibinfo{volume}{16}, \bibinfo{number}{2} (\bibinfo{year}{2008}), \bibinfo{pages}{467--476}.
\newblock
\urldef\tempurl%
\url{https://doi.org/10.1109/TASL.2007.913750}
\showDOI{\tempurl}


\bibitem[Wang et~al\mbox{.}(2024)]%
        {paper03}
\bibfield{author}{\bibinfo{person}{Jingyi Wang}, \bibinfo{person}{Alireza Sharifi}, \bibinfo{person}{Thippa Gadekallu}, {and} \bibinfo{person}{Achyut Shankar}.} \bibinfo{year}{2024}\natexlab{}.
\newblock \showarticletitle{{MMD-MII Model}: A multilayered analysis and multimodal integration interaction approach revolutionizing music emotion classification}.
\newblock \bibinfo{journal}{\emph{International Journal of Computational Intelligence Systems}}  \bibinfo{volume}{17} (\bibinfo{year}{2024}), \bibinfo{pages}{99}.
\newblock
\urldef\tempurl%
\url{https://doi.org/10.1007/s44196-024-00489-6}
\showDOI{\tempurl}


\bibitem[Wang et~al\mbox{.}(2014)]%
        {CAL500exp}
\bibfield{author}{\bibinfo{person}{Shuo-Yang Wang}, \bibinfo{person}{Ju-Chiang Wang}, \bibinfo{person}{Yi-Hsuan Yang}, {and} \bibinfo{person}{Hsin-Min Wang}.} \bibinfo{year}{2014}\natexlab{}.
\newblock \showarticletitle{Towards time-varying music auto-tagging based on {CAL500} expansion}. In \bibinfo{booktitle}{\emph{IEEE International Conference on Multimedia and Expo (ICME)}}. \bibinfo{publisher}{IEEE}, \bibinfo{address}{Chengdu, China}, \bibinfo{pages}{1--6}.
\newblock
\urldef\tempurl%
\url{https://doi.org/10.1109/ICME.2014.6890290}
\showDOI{\tempurl}


\bibitem[Wang et~al\mbox{.}(2011)]%
        {41_survey}
\bibfield{author}{\bibinfo{person}{Xing Wang}, \bibinfo{person}{Chen Xiaoou}, \bibinfo{person}{Deshun Yang}, {and} \bibinfo{person}{Yuqian Wu}.} \bibinfo{year}{2011}\natexlab{}.
\newblock \showarticletitle{Music emotion classification of {Chinese} songs based on lyrics using {TF*IDF} and rhyme}. In \bibinfo{booktitle}{\emph{International Society for Music Information Retrieval Conference (ISMIR)}}. \bibinfo{publisher}{International Society for Music Information Retrieval}, \bibinfo{address}{Miami, Florida, USA}, \bibinfo{pages}{765--770}.
\newblock
\urldef\tempurl%
\url{https://ismir2011.ismir.net/papers/PS6-19.pdf}
\showURL{%
\tempurl}


\bibitem[Wang and Guan(2008)]%
        {cnn_v_1}
\bibfield{author}{\bibinfo{person}{Yongjin Wang} {and} \bibinfo{person}{Ling Guan}.} \bibinfo{year}{2008}\natexlab{}.
\newblock \showarticletitle{Recognizing human emotional state from audiovisual signals}.
\newblock \bibinfo{journal}{\emph{IEEE Transactions on Multimedia}} \bibinfo{volume}{10}, \bibinfo{number}{5} (\bibinfo{year}{2008}), \bibinfo{pages}{936--946}.
\newblock
\urldef\tempurl%
\url{https://doi.org/10.1109/TMM.2008.927665}
\showDOI{\tempurl}


\bibitem[Wang et~al\mbox{.}(2000)]%
        {wang2000multimedia}
\bibfield{author}{\bibinfo{person}{Yao Wang}, \bibinfo{person}{Zhu Liu}, {and} \bibinfo{person}{Jin-Cheng Huang}.} \bibinfo{year}{2000}\natexlab{}.
\newblock \showarticletitle{Multimedia content analysis-using both audio and visual clues}.
\newblock \bibinfo{journal}{\emph{IEEE Signal Processing Magazine}} \bibinfo{volume}{17}, \bibinfo{number}{6} (\bibinfo{year}{2000}), \bibinfo{pages}{12--36}.
\newblock
\urldef\tempurl%
\url{https://doi.org/10.1109/79.888862}
\showDOI{\tempurl}


\bibitem[Watson et~al\mbox{.}(1999)]%
        {watson_T}
\bibfield{author}{\bibinfo{person}{David Watson}, \bibinfo{person}{David Wiese}, \bibinfo{person}{Jatin Vaidya}, {and} \bibinfo{person}{Auke Tellegen}.} \bibinfo{year}{1999}\natexlab{}.
\newblock \showarticletitle{The two general activation systems of affect: Structural findings, evolutionary considerations, and psychobiological evidence}.
\newblock \bibinfo{journal}{\emph{Journal of Personality and Social Psychology}} \bibinfo{volume}{76}, \bibinfo{number}{5} (\bibinfo{year}{1999}), \bibinfo{pages}{820--838}.
\newblock
\urldef\tempurl%
\url{https://doi.org/10.1037/0022-3514.76.5.820}
\showDOI{\tempurl}


\bibitem[Xuan et~al\mbox{.}(2020)]%
        {cross_graph_a_video_loc}
\bibfield{author}{\bibinfo{person}{Hanyu Xuan}, \bibinfo{person}{Zhenyu Zhang}, \bibinfo{person}{Shuo Chen}, \bibinfo{person}{Jian Yang}, {and} \bibinfo{person}{Yan Yan}.} \bibinfo{year}{2020}\natexlab{}.
\newblock \showarticletitle{Cross-modal attention network for temporal inconsistent audio-visual event localization}. In \bibinfo{booktitle}{\emph{AAAI Conference on Artificial Intelligence}}. \bibinfo{publisher}{AAAI Press}, \bibinfo{address}{Palo Alto, California, USA}, \bibinfo{pages}{279--286}.
\newblock
\urldef\tempurl%
\url{https://doi.org/10.1609/aaai.v34i01.5361}
\showDOI{\tempurl}


\bibitem[Xue et~al\mbox{.}(2015)]%
        {2015_1}
\bibfield{author}{\bibinfo{person}{Hao Xue}, \bibinfo{person}{Like Xue}, {and} \bibinfo{person}{Feng Su}.} \bibinfo{year}{2015}\natexlab{}.
\newblock \showarticletitle{Multimodal music mood classification by fusion of audio and lyrics}. In \bibinfo{booktitle}{\emph{International Conference on Multimedia Modeling (MMM)}}. \bibinfo{publisher}{Springer International Publishing}, \bibinfo{address}{Cham, Switzerland}, \bibinfo{pages}{26--37}.
\newblock
\urldef\tempurl%
\url{https://doi.org/10.1007/978-3-319-14442-9_3}
\showDOI{\tempurl}


\bibitem[Yang et~al\mbox{.}(2024)]%
        {paper04}
\bibfield{author}{\bibinfo{person}{Liang Yang}, \bibinfo{person}{Zhexu Shen}, \bibinfo{person}{Jingjie Zeng}, \bibinfo{person}{Xi Luo}, {and} \bibinfo{person}{Hongfei Lin}.} \bibinfo{year}{2024}\natexlab{}.
\newblock \showarticletitle{{COSMIC}: Music emotion recognition combining structure analysis and modal interaction}.
\newblock \bibinfo{journal}{\emph{Multimedia Tools and Applications}} \bibinfo{volume}{83}, \bibinfo{number}{5} (\bibinfo{year}{2024}), \bibinfo{pages}{12519--12534}.
\newblock
\urldef\tempurl%
\url{https://doi.org/10.1007/s11042-023-15376-z}
\showDOI{\tempurl}


\bibitem[Yang et~al\mbox{.}(2025)]%
        {paper25}
\bibfield{author}{\bibinfo{person}{Qi Yang}, \bibinfo{person}{Songhu Liu}, {and} \bibinfo{person}{Tianzhuo Gong}.} \bibinfo{year}{2025}\natexlab{}.
\newblock \showarticletitle{Improve the application of reinforcement learning and multi-modal information in music sentiment analysis}.
\newblock \bibinfo{journal}{\emph{Expert Systems}} \bibinfo{volume}{42}, \bibinfo{number}{1} (\bibinfo{year}{2025}), \bibinfo{pages}{e13416}.
\newblock
\urldef\tempurl%
\url{https://doi.org/10.1111/exsy.13416}
\showDOI{\tempurl}


\bibitem[Yang and Chen(2012)]%
        {2011-MER}
\bibfield{author}{\bibinfo{person}{Yi-Hsuan Yang} {and} \bibinfo{person}{Homer~H. Chen}.} \bibinfo{year}{2012}\natexlab{}.
\newblock \showarticletitle{Machine recognition of music emotion: A review}.
\newblock \bibinfo{journal}{\emph{ACM Transactions on Intelligent Systems and Technology}} \bibinfo{volume}{3}, \bibinfo{number}{3} (\bibinfo{year}{2012}), \bibinfo{pages}{40}.
\newblock
\showISSN{2157-6904}
\urldef\tempurl%
\url{https://doi.org/10.1145/2168752.2168754}
\showDOI{\tempurl}


\bibitem[Yang et~al\mbox{.}(2008)]%
        {2008_2}
\bibfield{author}{\bibinfo{person}{Yi-Hsuan Yang}, \bibinfo{person}{Yu-Ching Lin}, \bibinfo{person}{Heng-Tze Cheng}, \bibinfo{person}{I-Bin Liao}, \bibinfo{person}{Yeh-Chin Ho}, {and} \bibinfo{person}{Homer~H. Chen}.} \bibinfo{year}{2008}\natexlab{}.
\newblock \showarticletitle{Toward multi-modal music emotion classification}. In \bibinfo{booktitle}{\emph{Pacific-Rim Conference on Multimedia (PCM)}}. \bibinfo{publisher}{Springer}, \bibinfo{address}{Berlin, Germany}, \bibinfo{pages}{70--79}.
\newblock
\showISBNx{978-3-540-89796-5}
\urldef\tempurl%
\url{https://doi.org/10.1007/978-3-540-89796-5_8}
\showDOI{\tempurl}


\bibitem[Yin et~al\mbox{.}(2022)]%
        {EDA_music}
\bibfield{author}{\bibinfo{person}{Guanghao Yin}, \bibinfo{person}{Shouqian Sun}, \bibinfo{person}{Dian Yu}, \bibinfo{person}{Dejian Li}, {and} \bibinfo{person}{Kejun Zhang}.} \bibinfo{year}{2022}\natexlab{}.
\newblock \showarticletitle{A multimodal framework for large-scale emotion recognition by fusing music and electrodermal activity signals}.
\newblock \bibinfo{journal}{\emph{ACM Transactions on Multimedia Computing, Communications, and Applications}} \bibinfo{volume}{18}, \bibinfo{number}{3} (\bibinfo{year}{2022}), \bibinfo{pages}{78}.
\newblock
\showISSN{1551-6857}
\urldef\tempurl%
\url{https://doi.org/10.1145/3490686}
\showDOI{\tempurl}


\bibitem[Zentner et~al\mbox{.}(2008)]%
        {GEMS}
\bibfield{author}{\bibinfo{person}{Marcel Zentner}, \bibinfo{person}{Didier Grandjean}, {and} \bibinfo{person}{Klaus~R Scherer}.} \bibinfo{year}{2008}\natexlab{}.
\newblock \showarticletitle{Emotions evoked by the sound of music: Characterization, classification, and measurement}.
\newblock \bibinfo{journal}{\emph{Emotion}} \bibinfo{volume}{8}, \bibinfo{number}{4} (\bibinfo{year}{2008}), \bibinfo{pages}{494}.
\newblock
\urldef\tempurl%
\url{https://doi.org/10.1037/1528-3542.8.4.494}
\showDOI{\tempurl}


\bibitem[Zhang et~al\mbox{.}(2018b)]%
        {PMEmo}
\bibfield{author}{\bibinfo{person}{Kejun Zhang}, \bibinfo{person}{Hui Zhang}, \bibinfo{person}{Simeng Li}, \bibinfo{person}{Changyuan Yang}, {and} \bibinfo{person}{Lingyun Sun}.} \bibinfo{year}{2018}\natexlab{b}.
\newblock \showarticletitle{The {PMEmo} dataset for music emotion recognition}. In \bibinfo{booktitle}{\emph{International Conference on Multimedia Retrieval (ICMR)}}. \bibinfo{publisher}{Association for Computing Machinery}, \bibinfo{address}{New York, NY, USA}, \bibinfo{pages}{135–142}.
\newblock
\showISBNx{9781450350464}
\urldef\tempurl%
\url{https://doi.org/10.1145/3206025.3206037}
\showDOI{\tempurl}


\bibitem[Zhang and Tian(2022)]%
        {paper10}
\bibfield{author}{\bibinfo{person}{Lige Zhang} {and} \bibinfo{person}{Zhen Tian}.} \bibinfo{year}{2022}\natexlab{}.
\newblock \showarticletitle{Research on music emotional expression based on reinforcement learning and multimodal information}.
\newblock \bibinfo{journal}{\emph{Mobile Information Systems}} \bibinfo{volume}{2022}, \bibinfo{number}{1} (\bibinfo{year}{2022}), \bibinfo{pages}{2616220}.
\newblock
\urldef\tempurl%
\url{https://doi.org/10.1155/2022/2616220}
\showDOI{\tempurl}


\bibitem[Zhang et~al\mbox{.}(2022)]%
        {paper18}
\bibfield{author}{\bibinfo{person}{Meixian Zhang}, \bibinfo{person}{Yonghua Zhu}, \bibinfo{person}{Wenjun Zhang}, \bibinfo{person}{Yunwen Zhu}, {and} \bibinfo{person}{Tianyu Feng}.} \bibinfo{year}{2022}\natexlab{}.
\newblock \showarticletitle{Modularized composite attention network for continuous music emotion recognition}.
\newblock \bibinfo{journal}{\emph{Multimedia Tools and Applications}} \bibinfo{volume}{82}, \bibinfo{number}{5} (\bibinfo{year}{2022}), \bibinfo{pages}{7319–7341}.
\newblock
\showISSN{1380-7501}
\urldef\tempurl%
\url{https://doi.org/10.1007/s11042-022-13577-6}
\showDOI{\tempurl}


\bibitem[Zhang et~al\mbox{.}(2018a)]%
        {cnn_v_3}
\bibfield{author}{\bibinfo{person}{Shiqing Zhang}, \bibinfo{person}{Shiliang Zhang}, \bibinfo{person}{Tiejun Huang}, \bibinfo{person}{Wen Gao}, {and} \bibinfo{person}{Qi Tian}.} \bibinfo{year}{2018}\natexlab{a}.
\newblock \showarticletitle{Learning affective features with a hybrid deep model for audio-visual emotion recognition}.
\newblock \bibinfo{journal}{\emph{IEEE Transactions on Circuits and Systems for Video Technology}} \bibinfo{volume}{28}, \bibinfo{number}{10} (\bibinfo{year}{2018}), \bibinfo{pages}{3030--3043}.
\newblock
\urldef\tempurl%
\url{https://doi.org/10.1109/TCSVT.2017.2719043}
\showDOI{\tempurl}


\bibitem[Zhang et~al\mbox{.}(2021a)]%
        {features_N_1}
\bibfield{author}{\bibinfo{person}{Yong Zhang}, \bibinfo{person}{Cheng Cheng}, {and} \bibinfo{person}{Yidie Zhang}.} \bibinfo{year}{2021}\natexlab{a}.
\newblock \showarticletitle{Multimodal emotion recognition using a hierarchical fusion convolutional neural network}.
\newblock \bibinfo{journal}{\emph{IEEE Access}}  \bibinfo{volume}{9} (\bibinfo{year}{2021}), \bibinfo{pages}{7943--7951}.
\newblock
\urldef\tempurl%
\url{https://doi.org/10.1109/ACCESS.2021.3049516}
\showDOI{\tempurl}


\bibitem[Zhang et~al\mbox{.}(2021b)]%
        {cross_graph_video}
\bibfield{author}{\bibinfo{person}{Zongmeng Zhang}, \bibinfo{person}{Xianjing Han}, \bibinfo{person}{Xuemeng Song}, \bibinfo{person}{Yan Yan}, {and} \bibinfo{person}{Liqiang Nie}.} \bibinfo{year}{2021}\natexlab{b}.
\newblock \showarticletitle{Multi-modal interaction graph convolutional network for temporal language localization in videos}.
\newblock \bibinfo{journal}{\emph{IEEE Transactions on Image Processing}}  \bibinfo{volume}{30} (\bibinfo{year}{2021}), \bibinfo{pages}{8265--8277}.
\newblock
\urldef\tempurl%
\url{https://doi.org/10.1109/TIP.2021.3113791}
\showDOI{\tempurl}


\bibitem[Zhao et~al\mbox{.}(2022)]%
        {paper13}
\bibfield{author}{\bibinfo{person}{Jiahao Zhao}, \bibinfo{person}{Ganghui Ru}, \bibinfo{person}{Yi Yu}, \bibinfo{person}{Yulun Wu}, \bibinfo{person}{Dichucheng Li}, {and} \bibinfo{person}{Wei Li}.} \bibinfo{year}{2022}\natexlab{}.
\newblock \showarticletitle{Multimodal music emotion recognition with hierarchical cross-modal attention network}. In \bibinfo{booktitle}{\emph{IEEE International Conference on Multimedia and Expo (ICME)}}. \bibinfo{publisher}{IEEE}, \bibinfo{address}{Taipei, Taiwan}, \bibinfo{pages}{1--6}.
\newblock
\urldef\tempurl%
\url{https://doi.org/10.1109/ICME52920.2022.9859812}
\showDOI{\tempurl}


\bibitem[Zhao and Yoshii(2023)]%
        {paper11}
\bibfield{author}{\bibinfo{person}{Jiahao Zhao} {and} \bibinfo{person}{Kazuyoshi Yoshii}.} \bibinfo{year}{2023}\natexlab{}.
\newblock \showarticletitle{Multimodal multifaceted music emotion recognition based on self-attentive fusion of psychology-inspired symbolic and acoustic features}. In \bibinfo{booktitle}{\emph{Asia Pacific Signal and Information Processing Association Annual Summit and Conference (APSIPA ASC)}}. \bibinfo{publisher}{IEEE}, \bibinfo{address}{Taipei, Taiwan}, \bibinfo{pages}{1641--1645}.
\newblock
\urldef\tempurl%
\url{https://doi.org/10.1109/APSIPAASC58517.2023.10317539}
\showDOI{\tempurl}


\bibitem[Zhou et~al\mbox{.}(2019)]%
        {2019_01}
\bibfield{author}{\bibinfo{person}{Jianchao Zhou}, \bibinfo{person}{Xiaoou Chen}, {and} \bibinfo{person}{Deshun Yang}.} \bibinfo{year}{2019}\natexlab{}.
\newblock \showarticletitle{Multimodel music emotion recognition using unsupervised deep neural networks}. In \bibinfo{booktitle}{\emph{Conference on Sound and Music Technology (CSMT)}}. \bibinfo{publisher}{Springer}, \bibinfo{address}{Singapore}, \bibinfo{pages}{27--39}.
\newblock
\showISBNx{978-981-13-8707-4}
\urldef\tempurl%
\url{https://doi.org/10.1007/978-981-13-8707-4_3}
\showDOI{\tempurl}


\end{thebibliography}

% \appendix 
\end{document}